\documentclass[10pt,aps,pre,amsmath,amsfonts,amssymb,floatfix,superscriptaddress,notitlepage,nofootinbib]{revtex4}
\usepackage{mathrsfs} \usepackage{graphicx,color} \usepackage{bbm} \usepackage{multirow}
\usepackage{bm}

\usepackage{tikz}
\newcommand{\btkz}{\begin{tikzpicture}[baseline={([yshift=-.5ex]current bounding box.center)},vertex/.style={anchor=base}]}
\newcommand{\etkz}{\end{tikzpicture}}

\let\a=\alpha \let\b=\beta \let\g=\gamma \let\d=\delta
   
\let\l=\lambda    \let\p=\pi
\let\s=\sigma \let\t=\tau \let\f=\varphi 
   \let\G=\Gamma
\let\D=\Delta   
\let\Si=\Sigma  \let\Ps=\Psi 
 \let\r=\rho  \let\io=\infty

\def\FF{{\cal F}}

 \def\SS{{\cal S}}

\newcommand{\bx}{\mathbf{x}}
\newcommand{\bX}{\mathbf{X}}
\newcommand{\by}{\mathbf{y}}
\newcommand{\bY}{\mathbf{Y}}
\newcommand{\br}{\mathbf{r}}
\newcommand{\bk}{\mathbf{k}}
\newcommand{\bq}{\mathbf{q}}
\newcommand{\bu}{\mathbf{u}}
\newcommand{\bv}{\mathbf{v}}

\def\to{\rightarrow} \def\la{\left\langle} \def\ra{\right\rangle}

\newcommand{\beq}{\begin{equation}} \newcommand{\eeq}{\end{equation}}
\newcommand{\wh}{\widehat} 
\newcommand{\Tr}{\text{Tr}}

\begin{document}
\title{Quantitative approximation schemes for glasses}

\author{Matthieu Mangeat}
\affiliation{LPT,
\'Ecole Normale Sup\'erieure, UMR 8549 CNRS, 24 Rue Lhomond, 75005 Paris, France}
\affiliation{Master ICFP, D\'epartement de Physique, Ecole Normale Sup\'erieure, 24 Rue Lhomond,75005 Paris, France}

\author{Francesco Zamponi}
\affiliation{LPT,
\'Ecole Normale Sup\'erieure, UMR 8549 CNRS, 24 Rue Lhomond, 75005 Paris, France}

\begin{abstract}
By means of a systematic expansion around the infinite-dimensional solution, we obtain
an approximation scheme to compute properties of glasses in low dimensions. The resulting equations
take as input the thermodynamic and structural properties of the equilibrium liquid, and from this they allow
one to compute properties of the glass. They are therefore similar in spirit to the Mode-Coupling approximation
scheme. Our scheme becomes exact, by construction, in dimension $d\to\io$ and it
can be improved systematically by adding more terms in the expansion.
\end{abstract}

\maketitle

\tableofcontents

\clearpage

\section{Introduction}

\subsection{The exact solution of infinite-dimensional glassy hard spheres}
\label{sec:d_infty}

Many statistical mechanics models are exactly solvable in infinite spatial dimensions $d=\io$. 
The exact solution has a mean field structure,
and an expansion around this solution can be obtained in the form of a high temperature/low density expansion~\cite{GY91,FP99}.
Examples are the Ising model of magnetic materials~\cite{GY91}, which is described by the Curie-Weiss mean field theory for $d=\io$, and
a hard sphere liquid~\cite{FRW85,FP99}, which is described by the Van der Waals equation of state.

For the glass transition problem, one can apply the same strategy.
This approach was first proposed in~\cite{KW87}, but a complete and exact solution of the glass transition of identical hard spheres in $d=\io$ 
was obtained much more recently~\cite{PZ06a,PZ10,KPZ12,KPUZ13,CKPUZ13,CKPUZ14,RUYZ14,MKZ15}. 
The phase diagram turns out to be similar
to the one of a class of spin glass models (Ising $p$-spin and Potts glasses~\cite{GM84,Ga85,GKS85,KW87b,KT87,KT88,CC05}), thus confirming the 
main assumption behind the so-called 
Random First Order Transition (RFOT) theory of the glass transition~\cite{KT89,KTW89,WL12}. 

Within mean field (and thus RFOT theory),
the striking difference with respect to ordinary phase transitions is that the
glass phase is not a unique phase (like, e.g. a crystal). In the glassy region of the phase diagram, mutiple distinct glass states appear, each
characterized by different thermodynamic properties (e.g. a different equation of state).
The glassy phase diagram is thus extremely complex and characterized by several distinct phase transitions as a function of the control
parameters.
The exact solution allows one to compute the following list of physical quantities:
\begin{enumerate}
\item The full time-dependent correlations in the liquid phase, and in principle also the out-of-equilibrium correlations in the aging
regime~\cite{MKZ15}.
\item The dynamical  
transition density~\cite{KPUZ13,MKZ15}, at which the liquid phase becomes infinitely viscous and ergodicity is broken by
the emergence of many metastable states. This transition is similar to the one of Mode-Coupling Theory (MCT)~\cite{Go09} and
is thus characterized by MCT critical 
dynamical scaling, controlled by the so-called MCT parameter $\l$ that can be also computed~\cite{KPUZ13,MKZ15}.
\item The Kauzmann transition~\cite{PZ10}, where the number of metastable states becomes sub-exponential, giving rise
to an ``entropy crisis" and a second order equilibrium phase transition.
\item The Gardner transition line, that separates a region where glass basins are stable from a region where they are broken in a complex structure of metabasins~\cite{KPUZ13,CKPUZ14,RUYZ14}.
\item The density region where jammed packings exist (also known as ``jamming line'' or ``J-line''~\cite{MKK08,CBS09,PZ10}), 
which is delimited by the threshold density and the glass close packing density~\cite{PZ10}.
\item The equation of state of glassy states, computed by compression and decompression of equilibrium glasses~\cite{RUYZ14}.
\item The response of the glass state to a shear strain~\cite{YZ14,RUYZ14}.
\item The long time limit of the mean square displacement in the glass 
(the so-called Edwards-Anderson order parameter)~\cite{CKPUZ13}.
\item The behavior of correlation function, structural $g(r)$ and non-ergodicity factor of the glass~\cite{PZ10,CKPUZ13,MKZ15}.
\item The probability distribution of the forces in a packing, and the average number of particle contacts~\cite{CKPUZ13}.
\end{enumerate}
Note that the solution can also be extended to a wider class of potentials, including soft spheres and Lennard-Jones-like systems,
see~\cite{MKZ15, CKPUZ13}.

\subsection{Extension to finite dimension: state of the art}

The natural question once the $d=\io$ solution has been constructed is how to include finite dimensional corrections in a controlled way.
Here, one can explore two almost orthogonal research directions:
\begin{enumerate}
\item
The first problem is to include {\it quantitative} finite dimensional corrections.
In fact, in $d=\io$ typically only a few diagrams of the high temperature/low density expansion are relevant. In finite dimensions,
instead, all the diagrams contribute. Including a certain number (or even an infinite class) of diagrams typically does not change
the qualitative phase diagram of the system, which remains the same as in $d=\io$, but changes the quantitative results for
all the physical quantities (e.g. the transition temperature/density, or the specific heat). 
Away from the critical region around a phase transition, the inclusion of a few diagrams gives already quite good results
for the Ising model~\cite{GY91}. For particle systems, resummation of classes of diagrams are needed~\cite{Hansen}, leading 
to the successful approximations schemes of standard liquid theory, like the Hypernetted Chain (HNC) or Percus-Yevick (PY) approximations.
\item
The second problem is to study how finite dimensional corrections change the qualitative structure of the phase diagram. 
Typically, the most important qualitative changes are found in a rather small critical region around the phase transitions.
Below some upper critical dimension $d_u$ phase transitions change in nature, and the values of critical exponents change~\cite{ParisiBook}.
However, in some cases (typically in very low dimensions, below the lower critical dimension)
some phases or phase transitions that exist in $d=\io$ can be destroyed by finite-dimensional fluctuations. 
This problem has been very succesfully tackled by means of the renormalization group approach.
\end{enumerate}

Let us now discuss the specific case of glasses. 
The second problem is clearly the most interesting one from the intellectual point of view, and after the mean field theory was first proposed
by Kirkpatrick, Thirumalai and Wolynes~\cite{KT89,KTW89}, a sparkling debate has been centered around this issue.
In fact, the $d=\io$ phase diagram is dominated by infinitely-long lived {\it metastable} states, which become necessarily unstable through
nucleation in finite $d$~\cite{KTW89,BB04}. The lowest free energy, stable states (sometimes called {\it ideal glasses}),
lie in the region of the Kauzmann transition that is however completely inaccessible from the dynamical point of view~\cite{Ku01}.
Therefore, the dynamical transition cannot be more than a crossover in finite $d$~\cite{NBT15}, 
while the Kauzmann transition, if it exists, lie in a region
which is dynamically inaccessible. The finite dimensional laboratory glass transition happens in an intermediate region where metastability, nucleation, 
hopping, and critical fluctuations are all intertwined.
Nevertheless, some extremely interesting results have been obtained recently: for example, a field theory formulation of the dynamical
transition has been constructed, leading to the prediction of an upper critical dimension $d_u=8$~\cite{BB07,FPRR11} (but see also~\cite{NBT15})
and of the formulation of effective theories for $d<d_u$~\cite{FPRR11,FP13,Ri14,BCTT14,NBT15}. 
For the Kauzmann transition, renormalization group
studies indicate that it might survive in some cases and disappear in others~\cite{CDFMP10,CBTT11,YM12,CBTT13,AB14}.
Finally, instantonic techniques allowed to discuss nucleation in glasses~\cite{DSW05,Fr05,Fr07,FM07}.
Despite these partial successes, the situation remains confused, motivating alternative approaches that are based on phenomenological
models of constrained dynamics~\cite{Ga02,GST10}.

The first problem, in contrast, attracted somehow smaller attention. Still, from the practical point of view, it might be that all the subtle
effects that destroy the glass transition in finite dimensions happen on large time and length scales (see e.g.~\cite{NBT15}). In some cases, 
especially in numerical simulations and in experiments on mesoscopic or macroscopic particles (colloids and granulars), the accessible
time/length scales might be small enough that mean field theory remains a good description, provided one takes into account the quantitative
finite dimensional corrections. This point of view has been recently supported by extensive numerical simulations of hard spheres
in $d=13 \rightarrow 2$, that demonstrated that the evolution of the system is rather smooth as a function of $d$~\cite{CIPZ11,CIPZ12}.
This statement is especially true for static quantities, 
and deep in the dense/low temperature glass phase around the jamming transition~\cite{CCPZ12,GLN12}.
The results of~\cite{CIPZ11,CIPZ12,CCPZ12,GLN12} therefore motivate further development of an approximate mean field theory of glasses in finite~$d$.

The most known and most successful quantitative theory of glasses has historically been Mode-Coupling Theory (MCT)~\cite{Go09}.
MCT is a theory of dynamics, approaching the dynamical glass transition from the liquid phase. It belongs to the RFOT universality class,
because its qualitative predictions are identical to the ones of mean field theory~\cite{BCKM97,MKZ15}. 
It provides quantitatively good results in that region, in particular
for the non-ergodicity factor and for the shape of time-dependent correlations.
Despite its great successes, however, MCT also has some drawbacks.
First of all, the validity of MCT is mostly limited to the equilibrium liquid phase. Attemps to extend MCT to the glass phase have not been as
successful (see e.g.~\cite{IB13}). Second, MCT does not interpolate smoothly between $d=3$ and $d=\io$ and its predictions in large $d$
are not good~\cite{IM10,SS10,CIPZ11,MKZ15}. And finally, MCT is a dynamical theory, but we have seen that finite-dimensional processes
(in particular nucleation) that are neglected in MCT destroy the dynamical transition and deeply affect dynamics~\cite{NBT15}.

These observations motivated the development of a quantitative thermodynamic theory of glasses~\cite{MP96,CFP98,MP99,MP09}. 
In this setting one forgets about the dynamics and consider a {\it restricted} thermodynamic equilibrium to a metastable glass, 
and tries to describe the thermodynamic properties of the glass phase which are less affected by finite-dimensional corrections.
This approach has also been successful. It allows one to compute the equation of state and the specific heat of the glass, 
estimate the number of glassy states (the configurational entropy), and the Kauzmann temperature~\cite{MP96,MP99,MP09}.
Its extension to hard spheres leads to a unified phase diagram containing the glass and the jamming transitions~\cite{PZ10,BJZ11}
and allows one to compute some structural properties of the glass phase, in particular the correlation function $g(r)$. This approach
also allows one to compute the response of the glass to a shear strain~\cite{YM10,Yo12,Yo12b}.
The approximation schemes that are used~\cite{MP09,PZ10}, however, restrict the validity of this approach. In fact, the
{\it small cage expansion}~\cite{MP09,PZ10} holds only close enough to the Kauzmann point and does not allow one to make computations
close to the dynamical transition (e.g. the non-ergodicity factor is very poorly predicted~\cite{PZ10}). 
Moreover, computing the Gardner transition and the properties of the low-temperature Gardner phase
is very cumbersome within this approach~\cite{Ja2666}.

\subsection{Aim and structure of this paper}

The aim of this paper is to construct a simple approximation scheme that allows one to compute the same observables that have
been computed in the $d=\io$ solution of hard spheres, but for finite dimensional system. The scheme we will propose gives
exactly the same qualitative phase diagram of the $d=\io$ solution, with the same phase transitions and the same universal properties
(e.g. the critical exponents), but at the same time provides concrete numbers for all the physical quantities, that can be compared 
systematically with numerical simulations.
For the moment, our scheme is based on a static approach and on the use of replicas. However, its extension to dynamics seems
possible, following the strategy of~\cite{MKZ15}.

The resulting equations have the following structure: they take as ``input" the equation of state and the correlation function of the equilibrium
liquid (that can both be computed using liquid theory~\cite{Hansen}) and from that they give as ``output'' the replicated free energy
from which physical observables can be derived. This structure is very similar to MCT.
The results can be compared with the predictions of other approaches (e.g. MCT) and have comparable quality. However, the present
approach has some conceptual advantages:
\begin{enumerate}
\item Being based on an expansion around the $d=\io$ solution, the equations can be systematically improved by adding more terms to
the expansion.
\item The use of replicas allows one to compute all the quantities listed in section~\ref{sec:d_infty}, except the time-dependent correlators
(point 1). Note in particular that the dynamical transition and the MCT parameter $\l$ can be computed using replicas~\cite{CFLPRR12,KPUZ13}.
In particular, this approach can describe low-temperature glasses and the jamming transition.
\end{enumerate}

In the following we will first derive the general equations using a liquid theory approach. 
Next, we will show how previous approximation schemes
can be recovered as limiting cases of our approach.
Finally, we will extract from the equations some predictions
for the simplest quantities, and compare them with numerical simulations. 
This is done just to show that the predictions are reasonably good, but we do not attempt to perform
a precise comparison between theory and simulation.
Doing this, and extracting the complete phase diagram
(most notably the Gardner transition line) requires additional work, 
but we believe that the route is traced and the remaining quantities can be computed,
provided one has enough time (and computer time) to solve the equations numerically.

In the rest of this work, we adopt a more technical language and
we assume that the reader is familiar with liquid theory~\cite{Hansen},
with the general formulation of RFOT~\cite{CC05,Ca09,WL12},
with the replica method~\cite{Mo95,FP95} and its application to structural glasses~\cite{PZ10,MP09,BJZ11}
and with the perturbative methods discussed in~\cite{PZ05,PZ10} on which
the present work is strongly based.

\section{Derivation of the equations}
\label{sec:derivation}

We consider a system of $d$-dimensional classical point particles interacting through a pair potential $v(r)$, at temperature $T=1/\b$
(here $k_{\rm B}=1$).
A particle is described by a point $\bx$ in $d$ dimensions; in the following $x = |\bx|$. Typically for the same rotationally invariant
function we will use equivalently the notations $f(x)$ and $f(\bx)$.

\subsection{The variational problem}

Following Monasson~\cite{Mo95}, we consider $m$ identical replicas of the system; from the knowledge of their free energy
one can derive most of the physical observables in the liquid and glass phases~\cite{MP09,PZ10}. A complementary approach
is the Franz-Parisi potential (or ``state following'') method~\cite{FP95}: this only requires a minor modification of the replica structure,
that we do not explicitly discuss in the following, see~\cite{RUYZ14} for a discussion.
A ``molecule'' is made by $m$ identical copies of a given particle, as it is thus described by a point $\bar x = \{ \bx_1 \cdots \bx_m \}$. 
The starting point of our analysis is the expression of the free energy of a molecular liquid in terms of the single-particle density
and the pair correlation (see~\cite{Hansen,MH61} for the general diagrammatic formalism, and~\cite{MP99,PZ10} for its application
to the replicated problem):
\beq
\label{HNCfree}
\begin{split}
\b \Ps[\r(\bar x),g(\bar x,\bar y)] &= \frac{1}{2} \int d\bar x d\bar y \, \r(\bar x) \r(\bar y) 
\big[ g(\bar x,\bar y) \log g(\bar x,\bar y) 
 - g(\bar x,\bar y) + 1 + \b v(\bar x,\bar y) g(\bar x,\bar y) \big] \\
&+ \int d\bar x \r(\bar x) \big[ \log \r(\bar x) -1 \big]
+ \frac{1}{2} \sum_{n\geq 3} \frac{(-1)^n}{n} \Tr_x [ h\r]^n + \text{2PI diagrams} \ ,
\end{split}
\eeq
where, following standard liquid theory definitions~\cite{Hansen,MH61}:
\beq\label{eq:rhomol}
\r(\bar x) = \sum_{i=1}^N \la \d(\bar x - \bar x_i) \ra 
\eeq
 is the single-molecule density, $g(\bar x,\bar y)$ and $h(\bar x,\bar y)$ are the molecule-molecule pair distribution functions defined by
\beq\label{eq:rho2mol}
\begin{split}
\r^{(2)}(\bar x,\bar y) &= \sum_{i\neq j}^{1,N} \la \d(\bar x - \bar x_i)\d(\bar y - \bar x_j) \ra = \r(\bar x) \r(\bar y)g(\bar x,\bar y) \ , \\
h(\bar x,\bar y) &= g(\bar x,\bar y) - 1 \ , \\
\end{split} \eeq 
the interaction potential between two molecules
is $v(\bar x,\bar y)= \sum_a v(|\bx_a- \by_a|)$, and
\beq\begin{split}
\Tr_x [ h\r]^n &= \int d\bar x_1 \cdots d\bar x_n h(\bar x_1,\bar x_2) \r(\bar x_2) h(\bar x_2,\bar x_3) \r(\bar x_3) \cdots h(\bar x_{n-1},\bar x_n) \r(\bar x_n) h(\bar x_n,\bar x_1) \r(\bar x_1) \ .
\end{split}\eeq
``2PI diagrams'' indicates the class of 2-particle irreducible diagrams that we do not write explicitly, see~\cite{MH61} for details.
This expression is variational and should be minimized with respect to $\r$ and $g$. 
Following the strategy of~\cite{PZ05} (but using slightly different notations), we now make three approximations:
\begin{enumerate}
\item We neglect the 2PI diagrams and therefore consider the so-called HNC approximation~\cite{Hansen}. 
This is not strictly needed but it will simplify the calculations. The 2PI diagrams can be re-introduced systematically~\cite{PZ10}.

\item For the pair correlation, we assume a factorized form~\cite{PZ05}
\beq
\label{gprod}
g(\bar x,\bar y) = \prod_a G( \bx_a- \by_a )^{1/m} \ ,
\eeq
where $G(r)$ is an unknown function to be determined from the free energy minimization.

\item For the single-particle density, we assume the Gaussian form~\cite{MP99,MP09}
\beq
\label{rrho}
\r(\bar x) =\r \int d \bX \prod_a \g_A(\bx_a- \bX) \ , \hspace{30pt} 
\g_A(\bu)=\frac{e^{-\frac{u^2}{2A}}}{(2\p A)^{d/2}} \ .
\eeq
Here $\r$ is the number density of the non-replicated system.
Note that we thus have~\cite{PZ10,MP99,MP09}:
\beq\begin{split}
\frac{1}{N} \int & d\bar x \, \r(\bar x) \big[ \log \r(\bar x) -1 \big] =
\log \r -1 + \frac{d}{2} (1-m) \log ( 2\p A ) - \frac{d}{2} \log m + \frac{d}{2}(1-m) \ .
\end{split}\eeq
The parameter $A$, over which we will have to minimize the free energy, represents
the mean square displacement of particles due to vibrations in the glass phase~\cite{MP09,PZ10}.

\end{enumerate}

\subsection{``Small cage'' expansion}

We wish now to perform an expansion which is valid for small $A$, motivated by the observation that vibrations
in the glass are always quite small~\cite{MP09,PZ05,PZ10}.
In~\cite{PZ05} the expansion was organized by using
$A^{1/2}$ as the small parameter.
Here, instead, we try to avoid the development in powers of $A$. Our idea is that in all the integrals of the free energy we can write
\beq\label{eq:6}
g(\bar x, \bar y) = G(\bX-\bY) + \left[ \prod_a G(\bx_a- \by_a )^{1/m} - G(\bX-\bY) \right] \ ,
\eeq
and consider the contribution of the term in parenthesis to be small, because the Gaussian form of $\r(\bar x)$ forces the $\bx_a$ to be close to $\bX$ 
when $A$ is small.
For example, we have for the simplest term in the free energy
\beq\label{eq:7}
\begin{split}
&\frac{1}N \int d\bar x d\bar y \r(\bar x) \r(\bar y)g(\bar x, \bar y) \\
&= 
\frac{\r^2}N \int d\bX d\bY G(\bX-\bY) + \frac{\r^2}N \int d\bX d\bY \left[ \left( \int d\bu d\bv \g_A(\bu) \g_A(\bv) G(\bX-\bY+\bu-\bv)^{1/m} \right)^m - 
G(\bX-\bY) \right] \\
&= \r \int d\br G(\br) + \r \int d\br \left[ \left( \int d\bu \g_{2A}(\bu) G(\br - \bu)^{1/m} \right)^m - G(\br) \right] 
= \r \int d\br G(\br) + \r\int d\br Q(\br)
 \ ,
\end{split}\eeq
where we defined 
\beq
Q(r) = \left( \int d\bu \g_{2A}(\bu) G(\br - \bu)^{1/m} \right)^m - G(r) \ .
\eeq
In Eq.~\eqref{eq:7} we did not make any approximation, but in the following for more complicated integrals we will expand
assuming that the function $Q(r)$ is ``small''~\footnote{Note that $G(r)$ will turn out to coincide with the liquid correlation
function, and it is therefore a smooth function, with no structure on the small scale $A$}.

Let us now examine the ``ring diagrams'' (i.e. the terms $\Tr_x [h \r ]^n$) following~\cite{PZ05}. Defining $H(X) = G(X)-1$ 
and $C(X)$ from the Ornstein-Zernike relation~\cite{Hansen}
\beq\label{eq:OZ}
\r \int d \mathbf{Z} H( \bX- \mathbf Z) C( \mathbf Z- \bY) = H( \bX- \bY) - C( \bX- \bY) \ ,
\eeq
we have, expanding at first order in the expansion defined by Eq.~\eqref{eq:6}:
\beq
\begin{split}
\Tr_x[h \r ]^n = \r^n \Tr_X H^n + n \r^n \int d \bX_1 \cdots d \bX_n Q( \bX_1- \bX_2) H( \bX_2- \bX_3) \cdots H( \bX_n- \bX_1) \ ,
\end{split}
\eeq
and
\beq
\frac12 \sum_{n\geq 3} \frac{(-1)^n}{n} \Tr_x [ h\r]^n = \frac12 \sum_{n\geq 3} \frac{(-\r)^n}{n} \Tr_X H^n
- \frac{\r^2}{2} \int d \bX d \bY  Q( \bX- \bY) [ H( \bX- \bY) - C( \bX- \bY)]  \ .
\eeq

Finally we consider the remaining two terms in the free energy
\beq\begin{split}
&\frac1N \int d\bar x d\bar y \, \r(\bar x) \r(\bar y) g(\bar x,\bar y)
\big[  \log g(\bar x,\bar y) 
 + \b v(\bar x,\bar y) \big]
 = \frac{m}N \int d\bar x d\bar y \, \r(\bar x) \r(\bar y) g(\bar x,\bar y)
\left[  \frac1m \log G(\bx_1-\by_1) 
 + \b v(\bx_1 - \by_1) \right] \\
 &\sim \frac{m}N \int d\bar x d\bar y \, \r(\bar x) \r(\bar y) g(\bar x,\bar y)
\left[  \frac1m \log G(\bX-\bY) 
 + \b v(\bX - \bY) \right] \\
 &= \r \int d\br \, [ G(\br) + Q(\br) ]
\left[ \log G(\br) 
 + \b m v(\br) \right] \ ,
\end{split}\eeq
where we made an additional approximation by assuming that the arguments can be computed in $\bX-\bY$ instead of $\bx_1-\by_1$
at the lowest order\footnote{See~\cite{PZ10} for a discussion; an alternative is to use $\bx_1,\by_1$ instead of $\bX,\bY$ in Eq.~\eqref{eq:6}.}.

Collecting all terms together we obtain
\beq\label{eq:FF_HNC}
\begin{split}
\b \FF_m(A) =& \frac{\b \Psi}{N} = \b F_{\rm harm}(A) + \b m F_{\rm HNC}[G(r) ; \b m] + \b \D F[A,G(r)] \ , \\
\b m F_{\rm HNC} =& \frac{\r}{2} \int d \br \, \{ G(\br) \log G(\br) - G(\br) + 1 + \b m v(\br) G(\br) \} \\
&+ \frac{1}{2 \r} \int \frac{d \bk}{(2\p)^d} 
\left[ - \log[1+\r \hat H(\bk)] + \r \hat H(\bk) - \frac{\r^2}{2} \hat H(\bk)^2 \right] 
+ \log \r -1 \ , \\
\b F_{\rm harm} =& \frac{d}{2} (1-m) \log (2\p A) + \frac{d}{2} (1-m) 
- \frac{d}{2} \log m \ , \\
\b \D F =& 
 \frac\r2 \int d\br \,  Q(\br) 
\left[ \log G(\br) - 1
 + \b m v(\br) -H(\br) + C(\br) \right]
\end{split}
\eeq
where the Fourier transform has been defined as
\beq
\label{Fdef}
\hat H(\bk) = \int d\br \, e^{i \bk \br} H(\br) \ .
\eeq

\subsection{Determination of $G(r)$, and the final set of equations}

The minimization over $G(r)$ is now straightforward. In fact, at the lowest order, the correction $\D F$ does not contribute
and $G(r)$ is determined by minimization of $F_{\rm HNC}$.
We therefore obtain simply
\beq
G(r) = g_{\rm liq}(r ; T/m, \f) \ ,
\eeq
where $g_{\rm liq}$ is given by the HNC approximation, at the same density of the original problem and with the same
potential, but at a rescaled temperature $T/m$~\cite{MP99}.
Because the HNC equation
is $\log G(r) + \b m v(r) -H(r) + C(r) = 0$, we obtain our final result:
\beq\label{eq:FF_fin}
\begin{split}
\SS_m(A) = -\b \FF_m(A) =&- \b F_{\rm harm}(A) - \b m F_{\rm liq}(T/m,\f) - \b \D F \ , \\
\b F_{\rm harm} =& \, \frac{d}{2} (1-m) \log (2\p A) + \frac{d}{2} (1-m) 
- \frac{d}{2} \log m \ , \\
q_{m,A}(r)  =&  \int d\bu \g_{2A}(\bu)g_{\rm liq}(\br -\bu  ; T/m, \f)^{1/m} \ , \\
Q(r) =& \, q_{m,A}(r)^m - g_{\rm liq}(r ; T/m, \f) \ , \\
G_m(A) &= \frac{1}{V_d}\int d\br \,  Q(\br) =
d \int_0^\io dr r^{d-1} [q_{m,A}(r)^m - g_{\rm liq}(r ; T/m, \f) ] \ , \\
\b \D F =& 
- \frac\r2 \int d\br \,  Q(\br) = - 2^{d-1}\f G_m(A) \ . \\ 
\end{split}\eeq
Here 
\beq\label{eq:sfera}
\begin{split}
&V_d =\frac{\pi^{d/2}}{\G(1+d/2)} \ , \\
&\Omega_d = d \, V_d  =\frac{2 \pi^{d/2}}{\G(d/2)} \ , \\
&\f = \r V_d /2^d \ ,
\end{split}\eeq
are the volume of a $d$-dimensional unit sphere, and the $d$-dimensional solid angle, and the packing fraction
for hard spheres of unit diameter, respectively. $F_{\rm liq}(T/m,\f)$ is the equilibrium free energy of the liquid at temperature $T/m$ and density $\f$.
The resulting free energy $\FF_m(A)$ must be optimized over $A$ to complete the calculation.
Note that here the functions $\FF_m(A)$, $G_m(A)$, etc. obviously depend on all the control parameters,
but we do not indicate this dependence explicitly to keep a more compact notation.

\subsection{Discussion}

Although we derived them discarding the 2PI diagrams, and therefore within the HNC approximation,
Eqs.~\eqref{eq:FF_fin} manifestly require as input only the free energy of the liquid $F_{\rm liq}(T,\f)$
and its pair correlation function $g_{\rm liq}(r;T,\f)$. Given these quantities, one can compute $\FF_m(A)$ and
derive from it the thermodynamics and the structure of the glass.
The liquid quantities should be computed within the simple liquid
HNC approximation for consistency, but nothing prevents us from using other approximations, possibly more accurate,
as we will discuss in the following. This corresponds to a resummation of some class of the 2PI diagrams~\cite{Hansen}.

Also, here we only considered the simplest case where the problem has a replica symmetric (RS)
structure\footnote{Sometimes, for historical reason, the replica symmetric ansatz in the Monasson scheme is called
``1-step replica symmetry breaking". Here we call it replica symmetric because all replicas are manifestly equivalent.}, 
and all the replicas
are equivalent, as it is manifest in Eq.~\eqref{rrho}. A generalisation to a more complex replica symmetry breaking structure,
which is needed to describe low-temperature glasses and jamming~\cite{CKPUZ14},
is straightforward along the lines of the Gaussian derivation reported in Ref.~\cite{CKPUZ13}.

Our final set of Eqs.~\eqref{eq:FF_fin} can be thought as originating from a systematic expansion around the
infinite-dimensional limit. For this, let us first note that Eqs.~\eqref{eq:FF_fin}
coincide with the exact solution
of the problem in $d\to\io$ as discussed in~\cite{PZ10,KPZ12} (within the RS structure, as mentioned above). 
This can be easily shown analytically using the same steps as in~\cite[Appendix C.2]{PZ10} as follows:
\begin{itemize}
\item
the liquid correlation is $g_{\rm liq}(r ; T, \f) = e^{-\b v(r)}$ because only the first virial diagram survives~\cite{FRW85,FP99,PZ10};
\item
the convolution defining $q_{m,A}(r)$ can be computed using bipolar coordinates and reduces to a one-dimensional
integral that simplifies for $d\to\io$; the same happens for the function $G_m(A)$;
\item 
the free energy then becomes the one reported
 in~\cite[Section VI]{PZ10}\footnote{Recall that here we did not consider the general scheme that allows for replica symmetry breaking,
 which was discussed in~\cite{KPZ12,KPUZ13,CKPUZ13,CKPUZ14}.}.
\end{itemize} 
Finite-dimensional corrections originate from four different sources:
\begin{enumerate}
\item
the Gaussian form in Eq.~\eqref{rrho}, that leads to the exact result in $d\to\io$~\cite{KPZ12}, is only an approximation in finite $d$;
one should then add perturbatively corrections around the Gaussian ansatz;
\item
the product form in Eq.~\eqref{gprod} is also an approximation in finite $d$; again, one can add perturbatively the corrections
that mix different replicas;
\item 
in $d\to\io$, only the first virial term (two-body density interaction) matters~\cite{PZ10,KPZ12}, while in finite dimensions
all the virial coefficients contribute;
\item
for each virial term, one can make an expansion as in Eq.~\eqref{eq:6} and all orders contribute, leading to many-body
replica interactions~\cite[Appendix B]{PZ10}.
\end{enumerate}
In this paper, we completely discarded the corrections originating from items 1, 2, and 4; we only considered point 3
and resummed the contributions of the leading order in the expansion of Eq.~\eqref{eq:6} to all virial coefficients.
The result thus takes into account some, but not all, of the finite-$d$ corrections around the infinite-$d$ exact solution.
In principle, our equations can be improved by taking into account the other corrections systematically.

\section{Connections with previous work}

In this section, we show that from our final set of equations~\eqref{eq:FF_fin} one can derive, in special limits,
most of the replica approximation schemes that have been previously used.

\subsection{M\'ezard-Parisi 1999}

For smooth potentials, one can expand $G_m(A)$ in powers of $A$.
This small cage expansion of $G_m(A)$ can be performed along the lines of~\cite[Eq.(B4)]{BJZ11}, and it reads at the lowest order
\beq\begin{split}
G_m(A) &= - d A \frac{m-1}m \int_0^\io dr \, r^d g_{\rm liq}(r; T/m,\f) \D \left[  \log g_{\rm liq}(r; T/m,\f) \right] \ .
\end{split}\eeq
However, we do not recover exactly the result of the small cage expansion reported in~\cite{MP09}. 
To obtain it, we must expand
$\log g_{\rm liq}(r; T/m,\f) \sim -\b m v(r) + \cdots$, where $v(r)$ is the liquid potential, to get
\beq\begin{split}
G_m(A) =  d A \b (m-1) \int_0^\io dr \, r^d g_{\rm liq}(r; T/m,\f) \D v(r)  \ .
\end{split}\eeq
This is the result of~\cite{MP09}. It would be interesting to elucidate the reason of the difference between
these two results. 

\subsection{Parisi-Zamponi 2005}

Restricting to Hard Spheres and keeping the lowest order term in a $A^{1/2}$ expansion, we have~\cite{PZ05,PZ10}:
\beq
Q(r) = 2\sqrt{A} G(D) Q_m \d(r-D)  \ .
\eeq
Plugging this in Eq.~\eqref{eq:FF_HNC} or Eq.~\eqref{eq:FF_fin}
allows one to recover exactly the results of~\cite{PZ05,PZ10}.
The approximation of~\cite{PZ05,PZ10} thus consist in keeping only the lowest order in the $A^{1/2}$ expansion
of the function $G_m(A)$.

\subsection{Berthier-Jacquin-Zamponi 2011}

In~\cite{BJZ11} the following approximation was made for low-temperature soft spheres:
\beq
g_{\rm liq}(r; T, \f) \sim y_{\rm liq}^{\rm HS}(\f) e^{-\b v(r)} 
\ \ \ \Rightarrow \ \ \ g_{\rm liq}(r ; T/m, \f)^{1/m} = [y_{\rm liq}^{\rm HS}(\f)]^{1/m} e^{-\b v(r)}
\ .
\eeq
Plugging this in Eq.~\eqref{eq:FF_fin} gives back the expressions of~\cite{BJZ11}.

\section{Extracting physical quantities from the replicated free energy}

Here we describe how to extract physical quantities from the replicated free energy given in Eq.~\eqref{eq:FF_fin}.
We heavily rely on previous work where the physical justification of the recipes we use has been 
described~\cite{Mo95,MP09,PZ10,CC05}, and here we limit ourselves to give some details on how to apply these recipes.
We consider a generic potential $v(r)$, hence the control parameters are temperature $T$, packing fraction (or density)
$\f$, and the other parameters that appear in the potential (a sticky attraction, a long range repulsion, etc.).
These set of parameters define a multi-dimensional phase diagram in which the phase transition lines are defined.
For this reason below we refer generically to ``transition lines" and not to ``transition density", ``transition temperature'', etc.

\subsection{A convenient expression of $q_{m,A}(r)$}

First of all we need to obtain a more convenient expression for $q_{m,A}(r)$, which as expressed
in Eq.~\eqref{eq:FF_fin} is a $d$-dimensional convolution and therefore hard to compute numerically.
Using bipolar coordinates to compute the convolution~\cite{PZ10}, and recalling that
$I_n(x)$ is the modified Bessel function, we obtain an expression of this function in terms of a rather simple one-dimensional
integral:
\beq
q_{m,A}(r)=\int_0^\io du \, g_{\rm liq}(u  ; T/m, \f)^{1/m} \,
 \left( \frac{u}{r} \right)^{\frac{d-1}{2}} \frac{ e^{
-\frac{(r-u)^2}{4A} }}{\sqrt{4\p A}} \left[ e^{ -\frac{ru}{2A} } \sqrt{\pi \frac{ru}{A}} I_{ \frac{d-2}{2} } \left( \frac{ru}{2A} \right)\right] \ .
\eeq
Note that in particular for $m=1$ we have
\beq
q_{1,A}(r) = \int_0^\io du \, g_{\rm liq}(u; T,\f) \,
 \left( \frac{u}{r} \right)^{\frac{d-1}{2}} \frac{ e^{
-\frac{(r-u)^2}{4A} }}{\sqrt{4\p A}} \left[ e^{ -\frac{ru}{2A} } \sqrt{\pi \frac{ru}{A}} I_{ \frac{d-2}{2} } \left( \frac{ru}{2A} \right)\right]  \ ,
\eeq
and in the special case $d=3$ this expression further simplifies to
\beq\label{eq:qAd3}
q_{m,A}(r) = \int_0^\io du \, g_{\rm liq}(u; T/m,\f)^{1/m} \,
 \left( \frac{u}{r} \right) \frac{ e^{
-\frac{(r-u)^2}{4A} }}{\sqrt{4\p A}}
 \left[ 1 - e^{-\frac{r u}A} \right]  \ .
\eeq
Similar simplifications are obtained in all odd dimensions.

\subsection{The equation for $A$ and the complexity}

From Eqs.~\eqref{eq:FF_fin}, 
we can derive the equation for $A$ from the condition $\frac{\partial \SS_m(A)}{\partial A}=0$, which reads
\beq
\begin{split}
\label{eq:FA}
1&=\frac{2^d\varphi}{d} \frac{A}{1-m}\frac{\partial G_m(A)}{\partial A} =\frac{2^d\varphi}{d} F_m(A) \ , \\
F_m(A) &= \frac{dAm}{1-m} \int_0^\infty dr r^{d-1} \frac{\partial q_{m,A} (r)}{\partial A} q_{m,A}(r)^{m-1} \ .
\end{split}
\eeq
Note that $A$ is by itself a physical observable: it corresponds to the long time limit of the mean square displacement
in the glass phase.
For fixed $m$, $F_m(A)$ presents a maximum as a function of $A$.
For $\frac{2^d\varphi}{d} \max_A F_m(A) \leq 1$, Eq.~\eqref{eq:FA} has in general two solutions and the physical one
corresponds to the smallest value of $A$~\cite{PZ10}. 
Thus, the condition
\beq
\label{cluster}
1=\frac{2^d\varphi}{d} \max_A F_m(A) \ .
\eeq
which we call the ``RS spinodal",
separates the region of the phase diagram where Eq.~\eqref{eq:FA} has a non-trivial solution (the ``glass'') from the region
where there are no solutions (the ``liquid''). Note however that for $m<1$
the RS spinodal falls in a region which is unstable towards additional
steps of replica symmetry breaking~\cite{MR03,KPUZ13}.
For $m=1$, however, Eq.~\eqref{cluster} is stable and defines the dynamical glass transition, see below.

From Eqs.~\eqref{eq:FF_fin} we can derive the expression of the complexity~\cite{Mo95,PZ10} as follows
\beq
\label{sigma}
\begin{split}
\Sigma_m(A)&=\SS_m(A) - m \frac{\partial \SS_m(A)}{\partial m} =  S_{\rm liq} (T/m,\varphi) - \frac{d}{2} \log (2 \pi A) - d + \frac{d}{2} \log m + 2^{d-1} \varphi 
H_m(A) \ , \\
H_m(A)&=d\int_0^\infty dr r^{d-1}  \Big[q_{m,A}(r)^m - g_{\rm liq}(r; T/m,\varphi) \\
&-m^2 \frac{\partial q_{m,A}(r)}{\partial m} q_{m,A}^{m-1} (r) - m q_{m,A}(r)^m \log q_{m,A}(r) - \frac{T}{m}\partial_T g_{\rm liq}(r; T/m,\varphi)
\Big] \ .
\end{split}
\eeq
The complexity exists only if $A$ is a solution of Eq.~\eqref{eq:FA}, and a glassy state exists only if the complexity is positive. 
The ideal glass line is defined by the condition $\Sigma_m(A)=0$.

Note that the RS spinodal and the ideal glass line define the region in which a consistent RS glassy solution exists~\cite{PZ10}.

\subsection{The equilibrium transition densities: dynamical transition and Kauzmann transition}

The equilibrium line is defined by $m=1$~\cite{Mo95,MP09,PZ10,BJZ11}. 
The dynamical transition is the point where the solution for $A$ disappears (and thus the complexity also disappears) for $m=1$. It is
the solution of the condition
\beq
\label{phi_d}
1=\frac{2^d\varphi}{d} \max_A F_1(A) \ , \hskip30pt F_1(A) = - dA \int_0^\infty dr r^{d-1} \frac{\partial q_A (r)}{\partial A} \log q_{A}(r) \ ,
\eeq
where $q_A(r)$ is a shorthand notation for $q_{1,A}(r)$.
In the region where $1 \geq \frac{2^d\varphi}{d} \max_A F_1(A)$, the equation for $A$ has a solution.
In this case from Eq.~\eqref{sigma}, using the fact that $\int d\br q_{A} (\br) = \int d\br g_{\rm liq} (\br;T,\varphi)$, we get for the equilibrium complexity
\beq
\label{phi_k}
\begin{split}
\Sigma_{\rm eq} = \Si_1(A) &= S_{\rm liq} (\varphi) - \frac{d}{2} \log (2 \pi A) - d - 2^{d-1} d \varphi \int_0^\infty dr r^{d-1} 
[q_A(r) \log q_A(r) + \widetilde{q_A}(r) - T \partial_T g_{\rm liq}(r; T,\varphi)] 
 \ . \\
\end{split}
\eeq
Note that here $\Sigma_{\rm eq}$ is understood to be equal to $\Si_1(A)$ computed 
on the solution of $1 = \frac{2^d\varphi}{d} F_1(A)$ for $A$; therefore it does not depend on $A$.
The expression of $q_A(r) = q_{1,A}(r)$, $\frac{\partial q_A(r)}{\partial A}$ and $\widetilde{q_A}(r)$ are
\beq\label{eq:qAvarie}
\begin{split}
q_A(r)&= \int_0^\infty du g_{\rm liq}(u; T,\varphi) \left( \frac{u}{r} \right)^{\frac{d-1}{2}} \frac{e^{-\frac{(r-u)^2}{4A}}}{\sqrt{4\pi A}} \left[ e^{-\frac{ru}{2A}} \sqrt{\pi\frac{ru}{A}} I_{\frac{d-2}{2}}\left(\frac{ru}{2A}\right) \right] \ , \\
\frac{\partial q_A(r)}{\partial A}&= \int_0^\infty du g_{\rm liq}(u;T,\varphi) \left( \frac{u}{r} \right)^{\frac{d-1}{2}} \frac{e^{-\frac{(r-u)^2}{4A}}}{\sqrt{4\pi A}} \frac{1}{4A^2} \left\{ e^{-\frac{ru}{2A}} \sqrt{\pi\frac{ru}{A}} \left[ (r^2+u^2-2dA) I_{\frac{d-2}{2}}\left(\frac{ru}{2A}\right) -2 r u I_{\frac{d}{2}}\left(\frac{ru}{2A}\right) \right] \right\} \ , \\
\widetilde{q_A}(r) &= - \int_0^\infty du g_{\rm liq}(u,\varphi) \log  g_{\rm liq}(u;T,\varphi) \left( \frac{u}{r} \right)^{\frac{d-1}{2}} \frac{e^{-\frac{(r-u)^2}{4A}}}{\sqrt{4\pi A}} \left[ e^{-\frac{ru}{2A}} \sqrt{\pi\frac{ru}{A}} I_{\frac{d-2}{2}}\left(\frac{ru}{2A}\right) \right] \ .
\end{split}
\eeq
The Kauzmann transition is defined by the condition $\Si_{\rm eq} = 0$.

\subsection{The jamming line}
\label{sec:jamming}

The jamming transition, first systematically studied in~\cite{OSLN03}, happens when a system of particles with a hard core
cannot satisfy anymore all the hard core constraints. For densities below jamming, for $T\to 0$ one finds that $A$ remains finite,
while for densities above jamming, $A\to 0$ when $T\to 0$~\cite{PZ10,BJZ11}.
The condition to have a jamming line is thus that, in an appropriate limit, we have $g_{\rm liq}(r; T/m, \f)=0$ for $r \in [0,\s)$,
e.g. a hard core is induced in the system. Let us call $\t= T/m$.
There are three classes of potentials that can display a jamming line:
\begin{itemize}
\item Hard spheres (HS), with $v(r)=\io$ for $r<\s$ and $v(r)=0$ for $r\geq \s$. In this case $g_{\rm liq}(r; \f)$ does not depend on
$\t$ and vanishes obviously for $r \in [0,1)$.
\item Hard spheres with an additional potential (HS-P), i.e. $v(r)=\io$ for $r<\s$ and $v(r)$ is finite for $r\geq \s$. The difference with
the previous case is that $g_{\rm liq}(r; \t, \f)$ keeps its dependence on $\t$.
\item Soft spheres (SS) with $v(r) \geq 0$ for $r<\s$ and $v(r)=0$ for $r\geq \s$. In this case, we need to take the limit $\t\to 0$
to induce the hard core~\cite{BJZ11}, because $g_{\rm liq}(r; \t, \f)=0$ when $\t\to 0$ and $v(r) >0$. 
Note that the precise shape of $v(r)$ for $r<1$ becomes irrelevant when $\t\to0$: physically this corresponds to the fact that
the properties of jammed packings are independent of the details of the potential. 
\end{itemize}
Here we choose $\s=1$ for simplicity and without loss of generality\footnote{Note that this choice is consistent with the definition of packing
fraction $\f$ given in Eq.~\eqref{eq:sfera}.}. Moreover, below we do not indicate explicitly the dependence of $g_{\rm liq}(r; \t, \f)$ on $\t$:
for HS this is exact, for SS we have chosen implicitly $\t=0$. We do not discuss here the case of HS-P, where 
the dependence on $\t$ must be kept and an optimization
over $\t$ has to be done at the end to obtain the ground state~\cite{BJZ11}; the equations for this case can be easily derived
along the same lines.

The jamming limit corresponds to $m\to 0$ with $A=m \alpha$ and $\t = T/m$, followed by $\t\to0$ for SS.
Here we compute the complexity
in this limit.
Using $\lim_{z \rightarrow \infty} \sqrt{2\pi z} e^z I_n(z) =1 $ and $g_{\rm liq}(r;\varphi)=0$ for $r \in [0,1)$, we get
\beq\begin{split}
\lim_{m \rightarrow 0} q_{m,m\a}(r) &= \lim_{m \rightarrow 0} \int_1^\infty du g_{\rm liq}(u;\varphi)^{\frac{1}{m}} \left( \frac{u}{r} \right)^{\frac{d-1}{2}} \frac{e^{-\frac{(r-u)^2}{4 m \alpha}}}{\sqrt{4\pi m \alpha}} =
\begin{cases}
\underset{m \rightarrow 0}{\lim} g_{\rm liq}(r;\varphi)^{\frac{1}{m}} & r>1 \\
\underset{m \rightarrow 0}{\lim} g_{\rm liq}(1;\varphi)^{\frac{1}{m}} \left( \frac{1}{r} \right)^{\frac{d-1}{2}} \frac{e^{-\frac{(r-1)^2}{4 m \alpha}}}{\sqrt{4\pi m \alpha}} & r<1 \\
\end{cases} \\
\lim_{m \rightarrow 0} q_{m,m\a}(r)^m &= 
\begin{cases}
g_{\rm liq}(r;\varphi) & r>1 \\
g_{\rm liq}(1;\varphi) e^{-\frac{(r-1)^2}{4 \alpha}} & r<1 \\
\end{cases} \\
- \lim_{m \rightarrow 0} m^2 \frac{\partial q_{m,m\a}(r)}{\partial m} &=
\begin{cases}
\underset{m \rightarrow 0}{\lim} g_{\rm liq}(r;\varphi)^{\frac{1}{m}} \log g_{\rm liq}(r;\varphi) & r>1 \\
\underset{m \rightarrow 0}{\lim} g_{\rm liq}(1;\varphi)^{\frac{1}{m}} \log g_{\rm liq}(1;\varphi) \left( \frac{1}{r} \right)^{\frac{d-1}{2}} \frac{e^{-\frac{(r-1)^2}{4 m \alpha}}}{\sqrt{4\pi m \alpha}} & r<1 \\
\end{cases} \\
\end{split}\eeq
Defining $G_0(\a) = \lim_{m\to 0} G_m(m\a)$, and similarly
$F_0(\a)$ and $H_0(\a)$, and assuming that $\t \partial_\t g_{\rm liq}(r;\t, \varphi) \to 0$ for $\t\to 0$ so that we can discard the last term in the expression of $H_m(A)$ in Eq.~\eqref{sigma}, we obtain:
\beq
\begin{split}
G_0(\a) &= 
\lim_{m \rightarrow 0, A=\alpha m}
d \int_0^\io dr r^{d-1} [q_{m,A}(r)^m - g_{\rm liq}(r ; \f) ] =  d \,  g_{\rm liq}(1;\varphi) \int_0^1 dr r^{d-1} e^{-\frac{(r-1)^2}{4 \alpha}}  \ , \\
F_0(\alpha)&=\lim_{m \rightarrow 0, A=\alpha m} \frac{dA}{1-m} \frac{\partial}{\partial A} \int_0^\infty dr r^{d-1} [ q_{m,A}(r)^m - g_{\rm liq}(r;\varphi) ] = 
d \, \alpha \frac{\partial}{\partial \alpha} \int_0^1 dr r^{d-1} g_{\rm liq}(1;\varphi) e^{-\frac{(r-1)^2}{4 \alpha}} \\
&= \frac{d}{4 \alpha} g_{\rm liq}(1;\varphi) \int_0^1 dr r^{d-1} (r-1)^2 e^{-\frac{(r-1)^2}{4 \alpha}} \ , \\
H_0(\alpha)&=\lim_{m \rightarrow 0, A=\alpha m} d  \int_0^\infty dr r^{d-1} \left\{ q_{m,A}(r)^m \left[ 1 - \frac{ m^2}{q_{m,A}(r)} \frac{\partial q_{m,A}(r)}{\partial m} - \log  q_{m,A}(r)^m \right] - g_{\rm liq}(r;\varphi) \right\} \\
&= d  g_{\rm liq}(1;\varphi) \int_0^1 dr r^{d-1} e^{-\frac{(r-1)^2}{4 \alpha}} \left[1+ \frac{(r-1)^2}{4\alpha}\right] = G_0(\a) + F_0(\a)  \ .
\end{split}
\eeq
The equation for $\a$ is obtained from Eq.~\eqref{eq:FA} and is $1= \frac{2^d \varphi}{d}  F_0(\alpha)$. Using this, we obtain
from Eq.~\eqref{sigma} the jamming entropy $\Si_j = \lim_{m\to0} \Si_m(m \a)$:
\beq
\begin{split}
\Sigma_j &=  S_{\rm liq} (\varphi) - \frac{d}{2} \log (2 \pi \alpha) - d + 2^{d-1} \varphi H_0(\alpha,\varphi) 
 = S_{\rm liq} (\varphi)  - \frac{d}{2} [\log (2 \pi \alpha) +1] + 2^{d-1} \varphi G_0(\a) \ .
\end{split}
\eeq
As for $\Si_{\rm eq}$, $\Si_j$ is computed on the solution of the equation for $\a$ and thus it does not depend on $\a$.

$\Si_j$ gives the Edwards entropy of jammed packings~\cite{PZ10,BJZ11}, which thus exist when $\Si_j \geq 0$.
The lower density limit of existence of packings is defined by the point where the solution for $\a$ disappears; we call it
$\varphi_{\rm th}$ and it is defined by the condition
\beq
\label{phi_th}
\begin{split}
1= \frac{2^d \varphi}{d} \max_\alpha F_0(\alpha) \ . 
\end{split}
\eeq
Note that at least in the limit $d\to\io$ one can show that the RS computation substantially underestimates the threshold, because the solution
is unstable towards further steps of replica symmetry breaking~\cite{CKPUZ13}.
The upper density limit of existence of packings, which we call $\varphi_{\rm GCP}$~\cite{PZ10}, is given by the condition
$\Si_j=0$.

\subsection{Correlation functions and non-ergodicity factor}

To compute correlation functions, let us define the density field of replica $a$ and its correlation:
\beq
\r_a(\bx) = \sum_{i=1}^N \d(\bx - \bx_i^a) \ , \hskip30pt
\r_{ab}(\bx-\by) = \la \r_a(\bx) \r_b(\by) \ra = \sum_{ij}^{1,N} \la  \d(\bx - \bx_i^a) \d(\by - \bx_j^b) \ra \ .
\eeq
Due to replica symmetry, the only independent correlations are $\r_{aa}(\br) = \r_{11}(\br)$ for all $a$, and
$\r_{ab}(\br) = \r_{12}(\br)$ for all $a\neq b$.
The diagonal term gives the pair correlation function of the glass, while the off-diagonal part gives the non-ergodicity factor.
In fact, defining the time-dependent van Hove correlation~\cite{Hansen} $G(\bx-\by; t) = \la \r(\bx, t) \r(\by,0) \ra$, we have, in the glass phase,
$G(\br; 0) = \r_{11}(\br)$ and $G(\br; t\to\io) = \r_{12}(\br)$~\cite{PZ10,FJPUZ13}.
Moving to Fourier space we can define
 the dynamic structure factor $S(k,t)$ and 
the non-ergodicity factor $f(k)$ as
\beq\begin{split}
S(k,t) &\propto  \int d\br \, e^{i \bk \br} [ G(\br,t) - \r^2 ]= \langle \rho(\bk ,t) \rho(-\bk,0) \rangle - \r^2 (2\pi)^d \delta(\bk)  \ , \\
f(k) &=\frac{S(k,t\to\io)}{S(k,0)} 
=\frac{ \int d\br \, e^{i \bk \br} [\r_{12}(\br) - \r^2]
}{ \int d\br \, e^{i \bk \br} [\r_{11}(\br) - \r^2]
}
= \frac{ \hat\r_{12}(\bk)- \r^2 (2\pi)^d \delta(\bk)}{\hat\r_{11}(\bk)- \r^2 (2\pi)^d \delta(\bk)}
\ .
\end{split}\eeq

According to the definitions in Eqs.~\eqref{eq:rhomol} and \eqref{eq:rho2mol}, 
and using Eq.~\eqref{gprod} and \eqref{rrho},
we have
\beq\label{eq:r11r12}
\begin{split}
\r_{11}(\bx-\by) &=  
 \d(\bx - \by) \sum_{i=1}^N \la  \d(\bx - \bx_i^1) \ra +
\sum_{i \neq j}^{1,N} \la  \d(\bx - \bx_i^1) \d(\by - \bx_j^1) \ra 
\\
&= \r   \d(\bx - \by) 
 + \int d\bx_2\cdots d\bx_m d\by_2 \cdots d\by_m \rho^{(2)}(\bx,\bx_2,\cdots,\bx_m,\by,\by_2,\cdots,\by_m) \\
&= \rho \d(\bx - \by)
 + \rho^2 \int d\bx_2\cdots d\bx_m d\by_2 \cdots d\by_m
d\bX d\bY \prod_{a=1}^m [\gamma_A(\bx_a-\bX) \gamma_A(\by_a-\bY) g_{\rm liq}(\bx_a-\by_a)^{\frac{1}{m}}]_{\bx_1=\bx,\by_1=\by} \\
&=  \rho \d(\bx-\by)  
+ \rho^2 \int d\bX d\bY  \gamma_A(\bx-\bX) \gamma_A(\by-\bY) g_{\rm liq}(\bx-\by)^{\frac{1}{m}} q_{m,A}(\bX-\bY)^{m-1} \ , \\
\r_{12}(\bx-\by) &= \sum_{i=1}^N \la  \d(\bx - \bx_i^1) \d(\by - \bx_i^2) \ra +
\sum_{i \neq j}^{1,N} \la  \d(\bx - \bx_i^1) \d(\by - \bx_j^2) \ra \\
&= \int d\bx_3...d\bx_m \rho(\bx,\by,\bx_3,...,\bx_m) 
 + \int d\bx_2\cdots d\bx_m d\by_1 d\by_3 \cdots d\by_m \rho^{(2)}(\bx,\bx_2,\cdots,\bx_m,\by_1,\by,\by_3,\cdots,\by_m) \\
&= \rho \int d\bx_3...d\bx_m d\bX \prod_{a=1}^m \gamma_A(\bx_a-\bX) \\
& \ \  \ + \rho^2 \int d\bx_2\cdots d\bx_m d\by_1 d\by_3 \cdots d\by_m
d\bX d\bY \prod_{a=1}^m [\gamma_A(\bx_a-\bX) \gamma_A(\by_a-\bY) g_{\rm liq}(\bx_a-\by_a)^{\frac{1}{m}}]_{\bx_1=\bx,\by_2=\by} \\
&=  \rho \gamma_{2A}(\bx-\by)  
+ \rho^2 \int d\bX d\bY  \gamma_A(\bx-\bX)  \gamma_A(\by-\bY) q_{m,A/2}(\bx-\bY) 
 q_{m,A/2}(\bX-\by) q_{m,A}(\bX-\bY)^{m-2} \ .
\end{split}
\eeq

Let us now focus for simplicity on the equilibrium line, $m=1$. Using a compact diagrammatic expression of $\r_{12}$, we have
\beq\label{eq:diagr}
\begin{split}
\r_{11}(\bx-\by) &= \r \d(\bx - \by) + \r^2 g_{\rm liq}(\bx - \by) \ , \\
\rho_{12}(\bx-\by)&= \rho \gamma_{2A}(\bx-\by)+\rho^2  g_{12}(\bx-\by) \ ,
\hskip20pt
g_{12}(\bx-\by) =
\btkz[baseline={([yshift=-.5ex]current bounding box.center)},vertex/.style={anchor=base}]
\fill (0,-0.5) circle (2pt);
\draw (0,0.5) circle (2pt);
\draw (1.4,-0.5) circle (2pt);
\fill (1.4,0.5) circle (2pt);
\draw[dashed] (0,-0.5)--(1.4,0.5);
\draw (0,-0.5)--(0,0.5);
\draw (1.4,-0.5)--(1.4,0.5);
\draw (0,-0.5) sin (0.1,-0.34) cos (0.2,-0.5) sin (0.3,-0.66) cos (0.4,-0.5) sin (0.5,-0.34) cos (0.6,-0.5) sin (0.7,-0.66) cos (0.8,-0.5) sin (0.9,-0.34) cos (1,-0.5) sin (1.1,-0.66) cos (1.2,-0.5) sin (1.3,-0.34) cos (1.4,-0.5);
\draw (0,0.5) sin (0.1,0.34) cos (0.2,0.5) sin (0.3,0.66) cos (0.4,0.5) sin (0.5,0.34) cos (0.6,0.5) sin (0.7,0.66) cos (0.8,0.5) sin (0.9,0.34) cos (1,0.5) sin (1.1,0.66) cos (1.2,0.5) sin (1.3,0.34) cos (1.4,0.5); 
\node at (1.85,0) {$q_{A/2}$};
\node at (-0.35,0) {$q_{A/2}$};
\node at (1.05,-0.05) {$q_A^{-1}$};
\node at (0.7,0.8) {$\gamma_A$};
\node at (0.7,-0.8) {$\gamma_A$};
\node at (-0.1,0.8) {$\bx$};
\node at (1.4,0.9) {$\bX$};
\node at (-0.1,-0.8) {$\bY$};
\node at (1.4,-0.9) {$\by$};
\etkz \ .
\end{split}
\eeq
Note that $\r_{11}(\br) = G(\br,0)$ is the static pair correlation of the density field and it is the sum of two terms:
the delta term is the ``self'' contribution coming from the correlation of each particle with itself, while the second
term $g_{\rm liq}(\br)$ is the ``coherent'' contribution coming from correlations between distinct particles.
At large times in the glass phase, $\r_{12}(\br) = G(\br,t\to\io)$ contains two similar contributions: the first, Gaussian,
contribution $\gamma_{2A}(\br)$ describes the broadening of the self delta peak due to vibration of each particle in
its cage, while the second contribution $g_{12}(\br)$ describes the coherent part of the correlation. One can 
more conveniently write
\beq\label{eq:g12simple}
g_{12}(\br) = 
\int d\mathbf{u} d\mathbf{v}  \gamma_A(\mathbf{u})  \gamma_A(\mathbf{v}) 
\frac{q_{A/2}(\br + \mathbf{u} ) 
 q_{A/2}(\br + \mathbf{v} )}{ q_{A}(\br+ \mathbf{u}+ \mathbf{v}) } \  .
\eeq

In order to write these expression in Fourier space,
let us first recall that $g_{\rm liq}(\br)=1+ h_{\rm liq}(\br)$ and
therefore
\beq
\r^{-1} [\hat\r_{11}(\bq)- \r^2 (2\pi)^d \delta(\bq)] = 1 + \r \hat g_{\rm liq}(\bq)- \r (2\pi)^d \delta(\bq) 
=  1 + \r \hat h_{\rm liq}(q) =  S_{\rm liq}(q)
 \ .
\eeq
Similarly, we define $h_{12}(\br) = g_{12}(\br) -1$ and we have
\beq
\r^{-1} [\hat\r_{12}(\bq)- \r^2 (2\pi)^d \delta(\bq)] = e^{-A q^2} + \r \hat g_{12}(\bq)- \r (2\pi)^d \delta(\bq) 
=  e^{-A q^2} + \r \hat h_{12}(q) = S_{12}(q)
 \ .
\eeq
With these definitions, we have $f(q) = S_{12}(q)/S_{\rm liq}(q)$.

The task is then to compute $\hat h_{12}(q)$. Let us
recall that 
$\hat \gamma_A(k)=e^{-Ak^2/2}$,
define $\hat q_A^{-1}(\bk)$ as the Fourier transform of $1/q_A(\br)$, and
\beq
Q_A(k)=\hat q_A^{-1}(\bk) - (2\pi)^d \delta(\bk) = \int d\br \,  e^{i\bk\br} \,
 \frac{1-q_A(\br)}{q_A(\br)} \ .
 \eeq
Using the Feynman rules on the diagrammatic representation of $g_{12}(\br)$ in Eq.~\eqref{eq:diagr}, 
 we thus get in momentum space
 \beq\begin{split}
 \hat h_{12}(\bq) &= 
\hat g_{12}(\bq) - (2\pi)^d \delta(\bq) \\ & = - (2\pi)^d \delta(\bq)   +  \int \frac{d\bk_1}{(2\pi)^d} \frac{d\bk_2}{(2\pi)^d} \hat \gamma_A(\bk_1) \hat \gamma_A(\bk_2) \hat \gamma_A(\bq-\bk_1) \hat \gamma_A(\bq-\bk_2) \hat g_{\rm liq}(\bk_1) \hat g_{\rm liq}(\bk_2) \hat q_A^{-1}(\bq-\bk_1-\bk_2)  \\
&=  e^{-Aq^2} \Big[
2  \hat h_{\rm liq}(q) +  Q_A(q) 
+  \int \frac{d\bk_1}{(2\pi)^d} e^{-2 A [  \bk_1^2 -  \bq \bk_1]} \hat h_{\rm liq}(\bk_1)\hat h_{\rm liq}(\bq - \bk_1) \\
& \ \ \ + 2  \int \frac{d\bk_1}{(2\pi)^d} e^{-A [  \bk_1^2 -  \bq \bk_1]} \hat h_{\rm liq}(\bk_1) Q_A(\bq - \bk_1) \\
& \ \ \ +  \int \frac{d\bk_1}{(2\pi)^d} \frac{d\bk_2}{(2\pi)^d} e^{-A [  \bk_1^2 + \bk_2^2 -  \bq (\bk_1+ \bk_2)]}
\hat h_{\rm liq}(\bk_1) \hat h_{\rm liq}(\bk_2) Q_A(\bq-\bk_1-\bk_2) 
 \Big] \ .
\end{split}
\eeq
Using this expression is probably not simpler than computing $h_{12}(\br)$ in direct space and then performing a Fourier transform.
Finally, note that
\begin{enumerate}
\item When $A=0$, one has $q_A(r) = g_{\rm liq}(r)$ and $\g_A(\br) = \d(\br)$. It follows that $\r_{12}(r) = \r_{11}(r)$ and therefore
$f(q)=1$.
\item When $A\to \io$, one has $q_A(r) \to 1, \forall r$. It follows that $g_{12}(r)=1$ and $h_{12}(r)=0$, hence
$f(q) = e^{-A q^2}/S_{\rm liq}(q) \to 0, \forall q >0$.
\end{enumerate}
Using these properties it might be possible to prove that in general $0 \leq f(q) \leq 1$, as it should be on physical grounds: however,
we were unable to obtain the proof.

Note that from the general expression of $\r_{11}(\br)$ in Eq.~\eqref{eq:r11r12} one can deduce the pair correlation of the glass
when $m<1$, $g_{\rm G}(r) = [\r_{11}(\br) - \r \d(\br)]/\r^2$. In particular, in the limit $m\to0$, one can obtain the pair correlation
at jamming that displays interesting features~\cite{PZ10,CKPUZ13}. Because in the jamming limit the pair correlation around contact
 is dominated
by the contact value of $g_{\rm liq}(r)$, in the present approach one recovers the same results of~\cite{PZ10}. It would be interesting
to study the behavior of $g_{\rm G}(r)$ around $r=2$ where a jump and a singularity are known to appear around jamming.

\section{Numerical methods}
\label{sec:liquid}

\subsection{Integral equations of liquid theory}

In order to solve the equations of replica theory, we need to be able to compute $g_{\rm liq}(r; T,\varphi)$ and the entropy
$S_{\rm liq}(T,\f)$ or the free energy of the liquid. This can be done using standard methods of liquid theory~\cite{Hansen}.
In the following, we will present results in several dimensions $d$. For $d=3$, 
standard Picard iteration using three-dimensional radial Fourier transforms can be used to solve the problem. However, for
$d\neq 3$ it is useful to give some details on the method we used. 

Let us now focus on a simple liquid of hard spheres, where $g_{\rm liq}(r;\f)$ depends only on packing fraction $\f$,
and drop the suffix ``liq'' for simplicity in the rest of this section,
as well as the explicit indication of the control parameter $\f$, unless needed explicitly.
We define the function $\gamma(r)=h(r)-c(r)$,
recalling that $h(r) =g(r)-1$ and that $c(r)$ is defined by the Ornstein-Zernike relation, Eq.~\eqref{eq:OZ}.
We will use the HyperNetted Chain (HNC) and the Percus-Yevick (PY) closures of liquid theory~\cite{Hansen}.
The HNC equation can be written as
\beq\label{eq:HNC}
c(r)=e^{-\beta v(r)} e^{h(r)-c(r)}-1-[h(r)-c(r)] = e^{\gamma(r)-\beta v(r)} -1- \gamma(r)=
\begin{cases}
-1-\gamma(r) & r<1 \ , \\
e^{\gamma(r)}-1-\gamma(r) & r>1 \ ,
\end{cases}
\eeq
while the PY equation is
\beq\label{eq:PY}
c(r)=e^{-\beta v(r)} [1+ h(r)-c(r)]-1-[h(r)-c(r)] = e^{-\beta v(r)} [1+ \gamma(r)] -1- \gamma(r)=
\begin{cases}
-1-\gamma(r) & r<1 \ , \\
0 & r>1 \ .
\end{cases}
\eeq
The Ornstein-Zernike equation can be written easily in Fourier space
\beq\label{eq:OZf}
\hat \gamma(k)=\hat h(k)-\hat c(k)=\frac{\hat c(k)}{1-\rho \hat c(k)}-\hat c(k)=\frac{\rho \hat c(k)^2}{1-\rho \hat c(k)}
\eeq
Hence, provided we can discretize the Fourier transform appropriately, we can solve these equations by a simple iteration scheme
on $c(r)$ and $\g(r)$: we start from a guess for, say, $c(r)$; we compute $\g(r)$ using Eq.~\eqref{eq:OZf}; finally, we compute a new
guess for $c(r)$ using either Eq.~\eqref{eq:HNC} or Eq.~\eqref{eq:PY}. We repeat until convergence and then $g(r) = 1 + c(r) + \g(r)$.

Once $g(r)$ has been obtained, the free energy (or entropy) of the liquid is easily obtained in the HNC scheme from
Eq.~\eqref{eq:FF_HNC}. In the PY case it can be obtained by integrating the exact expression~\cite{Hansen}
\beq\label{eq:pS}
p= - \f \frac{dS}{d\f} = 1+2^{d-1} \varphi g(1;\varphi) 
\hskip20pt
\Rightarrow
\hskip20pt
S(\varphi)=1-\log \rho - 2^{d-1} \int_0^\varphi d\varphi' g(1;\varphi') \ .
\eeq

\subsection{Discrete Fourier transformation}

To solve these equation iteratively, we need to perform a discrete Fourier transformation in $d$ dimension. 
The strategy to do this in our code has been adapted from the code of Atsushi Ikeda~\cite{IM10}.
The function $c$ and $\gamma$ are radial: hence, we can perform the angular integration and the Fourier transform reduces to
a Hankel transformation defined (for any radial function $f$ which can represent $c$ or $\gamma$) by,
\beq
\hat f(k)=\frac{(2\pi)^{\frac{d}{2}}}{k^{\frac{d}{2}-1}} \int_0^\infty dr r^{\frac{d}{2}} J_{\frac{d}{2}-1}(kr) f(r) 
\ , \hskip30pt f(r)=\frac{(2\pi)^{-\frac{d}{2}}}{r^{\frac{d}{2}-1}} \int_0^\infty dk k^{\frac{d}{2}} J_{\frac{d}{2}-1}(kr) \hat f(k) \ .
\eeq
To simplify these equations, we can define new functions $F(r)=r^{\frac{d}{2}-1} f(r)$ and $\hat F(k)=k^{\frac{d}{2}-1} \hat f(k)$, we get thus
\beq
\label{Hankelcont}
\hat F(k)=(2\pi)^{\frac{d}{2}} \int_0^\infty dr r J_{\frac{d}{2}-1}(kr) F(r) \ , \hskip30pt F(r)=(2\pi)^{-\frac{d}{2}} \int_0^\infty dk k J_{\frac{d}{2}-1}(kr) \hat F(k) \ .
\eeq
The orthogonality of Hankel transformations
\beq
\int_0^\infty dr r J_{\frac{d}{2}-1}(kr) J_{\frac{d}{2}-1}(k'r) = \frac{\delta(k-k')}{k}
\eeq
permits to assure the consistency of the previous equations.

We need now to discretize these equations on a grid of $N$ elements with $r$ in an interval $[0,R_{max}]$ and $k$ in an interval $[0,K_{max}]$ to solve the problem numerically. But, we also need to preserve the orthogonality property. A good way to do this is to cut the intervals in the zeros of the $(\frac{d}{2}-1)^{th}$ order Bessel function. We call $\lambda_i$ the $i^{th}$ zero of $J_{\frac{d}{2}-1}$, such that $\lambda_i \ne 0$. We take thus $r_i=\frac{\lambda_i}{K_{max}}$ and $k_i=\frac{\lambda_i}{R_{max}}$. $R_{max}$ and $K_{max}$ are thus related as $R_{max}\cdot K_{max}=\lambda_N$.
The equations (\ref{Hankelcont}) become
\beq
\label{Hankeldisc}
\begin{split}
\hat F(k_i)&= \sum_{j=1}^N (RK)_{ij} F(r_j) \ , \hskip30pt (RK)_{ij}= \frac{2(2\pi)^{\frac{d}{2}}}{K_{max}^2} \frac{J_{\frac{d}{2}-1}(k_i r_j)}{J_{\frac{d}{2}}(K_{max} r_j)^2}  \ , \\
F(r_i)&= \sum_{j=1}^N (KR)_{ij} \hat F(r_j) \ , \hskip30pt (KR)_{ij}= \frac{2(2\pi)^{-\frac{d}{2}}}{R_{max}^2} \frac{J_{\frac{d}{2}-1}(k_j r_i)}{J_{\frac{d}{2}}(R_{max} k_j)^2} \ .
\end{split}
\eeq
When $N$ is large, the continuous and the discrete version are equivalent, using the asymptotic expressions 
$J_{\frac{d}{2}}(x)= \sqrt{\frac{2}{\pi x}} \cos(x-(d-1)\frac{\pi}{4})$ and $\lambda_i=(i+\frac{d-3}{4})\pi$,
from which
\beq
 J_{\frac{d}{2}}(K_{max} r_j)^2 \sim \frac{2}{\pi K_{max} r_i} \text{~~and~~} J_{\frac{d}{2}}(R_{max} k_j)^2 \sim \frac{2}{\pi R_{max} k_j} \ .
\eeq
From this, we get the correct large $N$ limit:
\beq
\begin{split}
\hat F(k_i) &\sim \frac{\pi(2\pi)^{\frac{d}{2}}}{K_{max}} \sum_{j=1}^N r_j F(r_j) J_{\frac{d}{2}-1}(k_i r_j) \sim (2\pi)^{\frac{d}{2}} \int_0^{R_{max}} dr r J_{\frac{d}{2}-1}(k_j r) F(r) \ , \\
F(r_i) &\sim \frac{\pi(2\pi)^{-\frac{d}{2}}}{R_{max}}\sum_{j=1}^N k_j \hat F(k_j) J_{\frac{d}{2}-1}(k_j r_i) \sim (2\pi)^{-\frac{d}{2}} \int_0^{K_{max}} dk k J_{\frac{d}{2}-1}(kr_i) \hat F(k) \ ,
\end{split} 
\eeq
and the consistency property, using $x_i=\lambda_i/\lambda_N$
\beq
\begin{split}
F(r_i) &\sim \frac{\pi^2}{\lambda_N} \sum_{j,l} r_l k_j F(r_l) J_{\frac{d}{2}-1}(k_l r_j) J_{\frac{d}{2}-1}(k_j r_i) = \pi^2 \sum_{j,l} \lambda_l x_j F(r_l) J_{\frac{d}{2}-1}(\lambda_l x_j) J_{\frac{d}{2}-1}(\lambda_i x_j) \\
&\sim \sum_l F(r_l) \lambda_l \int_0^1 dx x J_{\frac{d}{2}-1}(\lambda_l x) J_{\frac{d}{2}-1}(\lambda_i x) = \sum_l F(r_l) \delta_{il} \ .
\end{split}
\eeq

\subsection{Algorithm for $g_{\rm liq}$}

In summary, to solve the integral equations of liquid theory, the following iterative algorithm is applied :
\begin{enumerate}
\item Give some arguments : $d$, $N$, $R_{max}$ and $\rho$.
\item Evaluate \{$\lambda_i$\}, $K_{max}$, \{$r_i$\}, \{$k_i$\}, $(RK)$ and $(KR)$ and keep them in memory (this step must be done only once
at the beginning, because these quantities depend on $d$, $N$, $R_{max}$ only).
\item Take an initial form of $\gamma$ : here we use $\gamma(r)=0$.
\item Do a recursive sequence to evaluate $\gamma$ and $c$ :
\begin{enumerate}
\item Evaluate $c$ from HNC or PY equation.
\item Evaluate $C(r_i)=r_i^{\frac{d}{2}-1} c(r_i)$, $\hat C (k_i)$ with $(RK)$ from (\ref{Hankeldisc}) and $\hat c(k_i) = k_i^{1-\frac{d}{2}} \hat C(k_i)$.
\item Evaluate $\hat \gamma$ with OZ equation.
\item Evaluate $\hat \Gamma(k_i)=k_i^{\frac{d}{2}-1} \hat \gamma(k_i)$, $\Gamma (r_i)$ with $(KR)$ from (\ref{Hankeldisc}) and $\gamma(r_i) = r_i^{1-\frac{d}{2}} \Gamma(r_i)$.
\item Guess a new value of $\gamma$ from $\gamma(r)=(1-\alpha)\gamma_{old}(r)+\alpha \gamma_{new}(r)$, for $\alpha$ small enough in order to have a good convergence of algorithm. A faster option is to use the DIIS algorithm\footnote{See e.g.~\url{https://en.wikipedia.org/wiki/DIIS} and references therein.} which uses the value of $\gamma$ in the last $n$ steps to speed up convergence.
\item The iteration stops when $|\gamma_{old}-\gamma_{new}|<10^{-10}$.
\end{enumerate}
\item Return $g(r)=\gamma(r)+c(r)+1$ as a linear interpolation of its discretized values, with a precaution concerning the discontinuity in $r=1$: we impose $g(r<0)=0$ and $g(1^+)$ is evaluated by interpolating linearly the first two values of $r>1$.
\end{enumerate}
As an example of the efficiency of this algorithm, for a $3.5 \text{~GHz}$ CPU computer, 
the initialization step costs about 30 seconds and the converging sequence needs 10 seconds for a single value of density and
for $N=3000$, using a C++ code with the GSL library to compute Bessel functions.

\subsection{More accurate expressions for three-dimensional hard spheres}

In the special case of three-dimensional hard spheres, very accurate expressions exist for the liquid entropy and pair correlation~\cite{Hansen}.
For $S_{\rm liq}$, one can use the Carnahan-Starling (CS) expression
\beq
\label{CSEQ}
 g_{\rm CS}(1;\varphi)=\frac{1- \frac12 \varphi}{(1-\varphi)^3} \ ,
\eeq
which, together with Eq.~\eqref{eq:pS}, gives an approximation of the liquid entropy $S(\f)$.
From the CS expression, Verlet and Weis derived an accurate fitting form of $g(r)$~\cite{VW72,Hansen}.
It is obtained as follows.
The PY result is modified by an additive function and by a phase shift:
\beq
\label{VWEQ}
g_{\rm VW}(r;\varphi)=g_{\rm PY}(\xi r;\varphi^*)+\delta g(r) \  ,
\hskip30pt
\delta g(r) = \frac{\delta g_1}{r} \cos[\alpha (r-1)] e^{-\alpha(r-1)} \ .
\eeq
Here $\d g_1$ is chosen to have $g_{\rm VW}(1)=g_{\rm CS}(1)$, given by (\ref{CSEQ}),
$\varphi^*$ is chosen to achieve a minimum in the absolute difference between the exact $g(r)$ (obtained by numerical simulations) 
and $g_{\rm VW}$, 
and $\alpha$ to obtain consistently the isothermal compressibility derived from the CS expression of the entropy.
Explicitly, the parameters are 
\beq
\varphi^*=\varphi-\frac{\varphi^2}{16} \ , \hskip20pt
\xi=\left(\frac{\varphi}{\varphi^*}\right)^{1/3} \ , \hskip20pt
\delta g_1=g_{\rm CS}(1;\f)-g_{\rm PY}(\xi;\f^*) \ , \hskip20pt
\alpha=\frac{24 \delta g_1}{\varphi^* g_{\rm PY}(1;\f^*)} \ .
\eeq 
This result fits the exact computer-generated function with an error less than 1\% for all $\varphi$.
Thus, for the Verlet-Weis (VW) approximation, the numerical algorithm is applied to find $g_{\rm PY}$ at $\varphi^*$ 
and the pair correlation function is then obtained from Eq.~(\ref{VWEQ}).

\subsection{Non-ergodicity factor}
\label{sec:fqnum}

The calculation of the non-ergodicity factor is particularly cumbersome and here we specialize to $d=3$.
This calculation takes as input the values
of $g_{\rm liq}(r)$ and $A$.
Then, one can compute and tabulate $q_A(r)$, which is a simple one-dimensional integral given by Eq.~\eqref{eq:qAd3}.
Next, we found simpler to use Eq.~\eqref{eq:g12simple} to compute $g_{12}(\br)$: this is done by constructing
a three-dimensional grid for $\mathbf{u}$ and $\mathbf{v}$. Finally, we take the Fourier transform of the result
to obtain~$f(q)$.

Unfortunately, this procedure suffers from large numerical inaccuracies at small $q$. This is due to the singularity
of $g_{\rm liq}(r)$ at $r=1$; for small $A$, a correct integration of this singularity requires a very small step of the
grid in $r$. And even then, because the values of $S_{\rm liq}(q)$ and $S_{12}(q)$ are both quite small for $q\to 0$,
even a small error can affect heavily the determination of $f(q)$. In the end, we were not able to obtain reliable
results for $q \lesssim 5$ (in units of the sphere diameter).

\subsection{Summary}

In summary, from the procedures described above we can obtain the HNC, PY or VW expressions for the entropy $S_{\rm liq}(\f)$
and the pair correlation $g_{\rm liq}(r;\f)$ of hard spheres in all dimensions (but only in $d=3$ for VW).
For other potentials, we will report briefly some results obtained from the HNC approximation in $d=3$ only.

Once the liquid quantities have been computed, one has to solve the equations for the various transition densities and for the complexity.
As an example, to obtain $\f_{\rm d}$ from Eqs.~\eqref{phi_d}, \eqref{eq:qAvarie}, one has to
\begin{enumerate}
\item 
Perform one-dimensional integrals to compute
$q_A(r)$ and $\partial q_A(r)/\partial A$ according to Eq.~\eqref{eq:qAvarie}; this can be done using the same grid we used for the Fourier
transforms, or interpolating on a regular grid.
\item
Use the result to compute $F_1(A)$ according to Eq.~\eqref{phi_d}; this can be done again using integration
over the same grid. 
\item
Find the maximum of $F_1(A)$ over $A$; this is done using standard bisection methods.
\end{enumerate}
For each of these steps we actually used additional tricks to improve the convergence and the efficiency of the algorithm: but none of those
are crucial, so we do not discuss them here. The reader will be able to find a way to perform these steps and reproduce our results without too much difficulty.

\section{Results: hard spheres}

We will now present some results that can be derived from this approach. Let us recall that in order to derive precise predictions
from a given interaction potential $v(r)$ in a given spatial dimension $d$, one should first solve the problem of computing
the liquid quantities, the entropy $S_{\rm liq}$ and the pair correlation $g_{\rm liq}$. These quantities are used as input
for the replica equations of Sec.~\ref{sec:derivation}, from which one can derive all the properties of the glass. Thus, in order
to obtain reliable predictions, it is quite important that accurate estimates of liquid quantities are available. This can be done
following the methods of~\cite{Hansen}, but no general recipe is known: for each interaction potential, one has to choose an
approximation that provides good result, and test it against numerical data for the liquid quantities. Another option is to use
as input for the replica equations the numerically, or experimentally, measured (and fitted) quantities. All this requires a preliminary
assessment of the accuracy of the description of the liquid.

Here we are not interested in obtaining the most precise predictions for a given potential. Our aim is to show that this approach
provides reasonably good numbers, and that in the limit $d\to\io$ the exact solution 
of~\cite{PZ06a,PZ10,KPZ12,KPUZ13,CKPUZ13,CKPUZ14,RUYZ14,MKZ15} is recovered.
We will thus focus on hard spheres in $d$ dimensions
and we will use the simplest approximation schemes of liquid theory, namely the HNC and PY
equations described in Sec.~\ref{sec:liquid}. Note that their accuracy in $d>3$ at high density close to the glass transition is not known. 
In $d=3$ they perform reasonably well but with evident discrepancies~\cite{Hansen}: in that special case
we will use, always for the purpose of illustrating the method, the more accurate Verlet-Weis (VW) approximation.
We will not report a detailed comparison of our results with existing numerical and experimental data, once again
because our aim here is to illustrate the main properties of this approach.

\subsection{HNC and PY results as a function of dimension}

\begin{figure}[t]
\includegraphics[width=.45\textwidth]{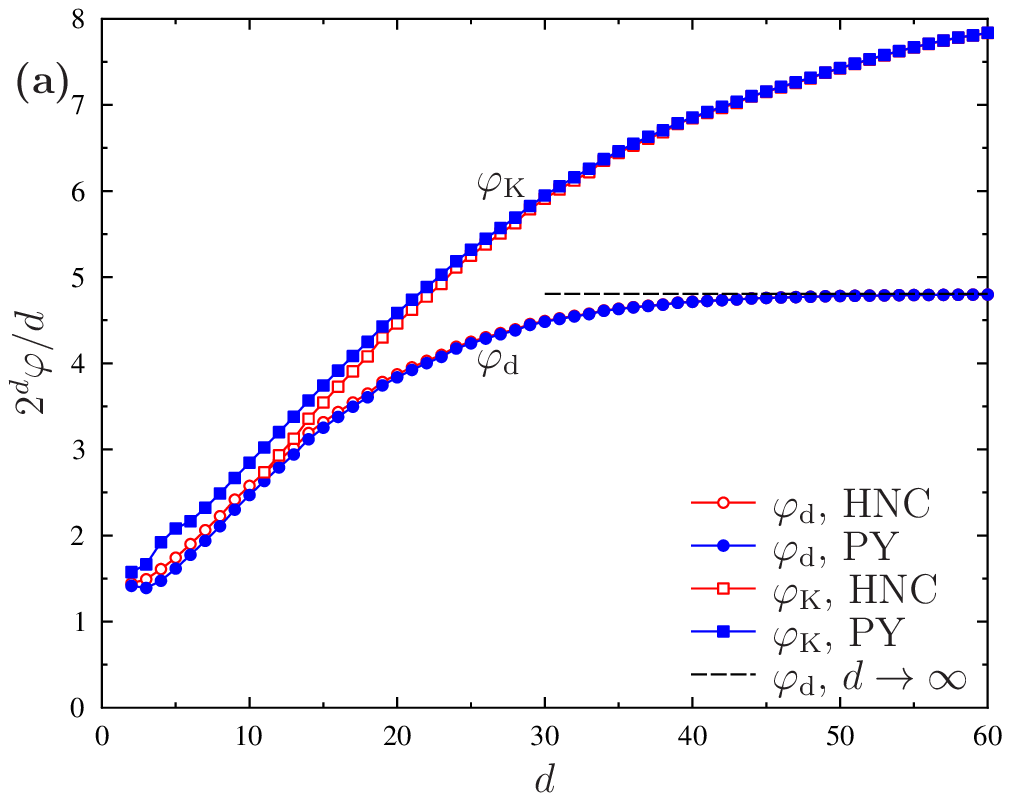}
\includegraphics[width=.45\textwidth]{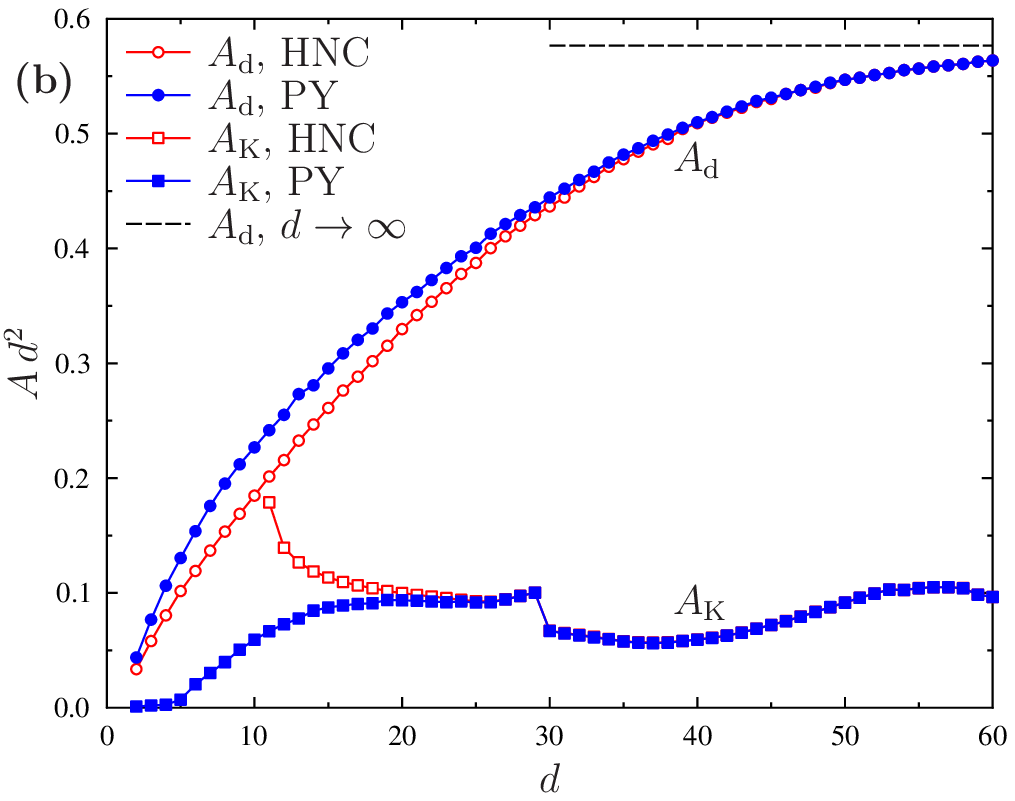}
\caption{(a)~Rescaled dynamical $\f_{\rm d}$ and Kauzmann $\f_{\rm K}$ packing fractions as a function of spatial dimension $d$.
(b)~Rescaled values of the cage radius $\wh A = A \, d^2$ at the dynamical and Kauzmann points. The jump in $A_{\rm K}$ that is
observed around $d=30$ is due to a numerical issue, see text.
}
\label{fig:1}
\end{figure}

\begin{figure}[t]
\includegraphics[width=.45\textwidth]{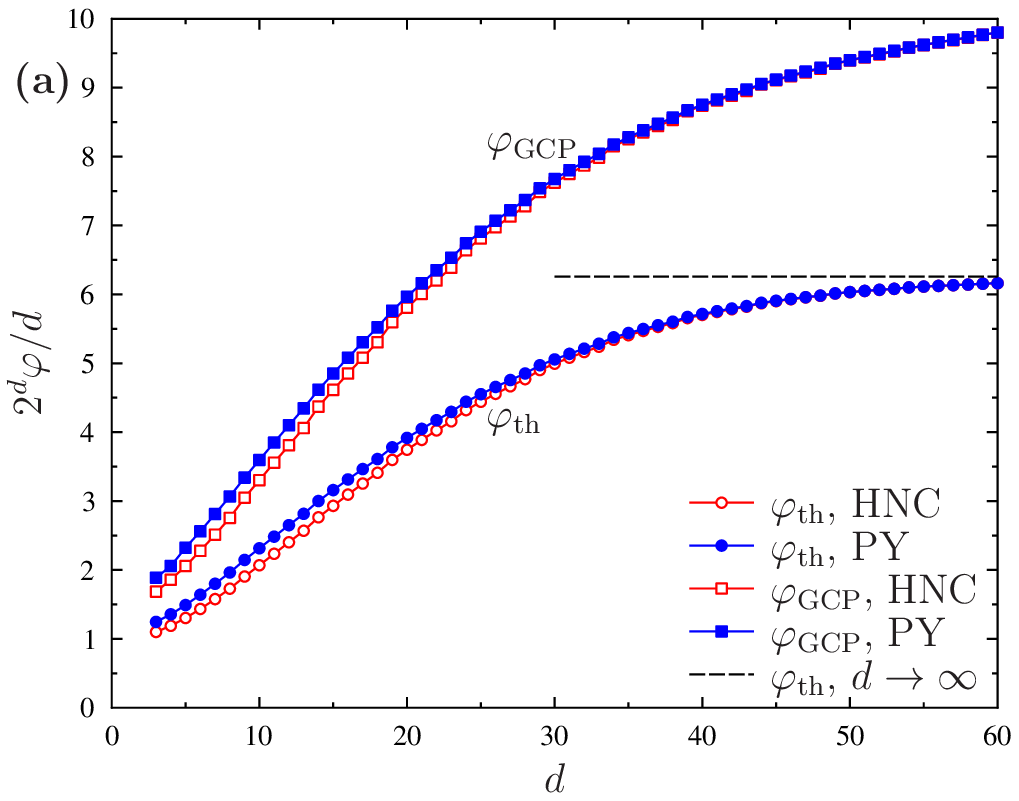}
\includegraphics[width=.45\textwidth]{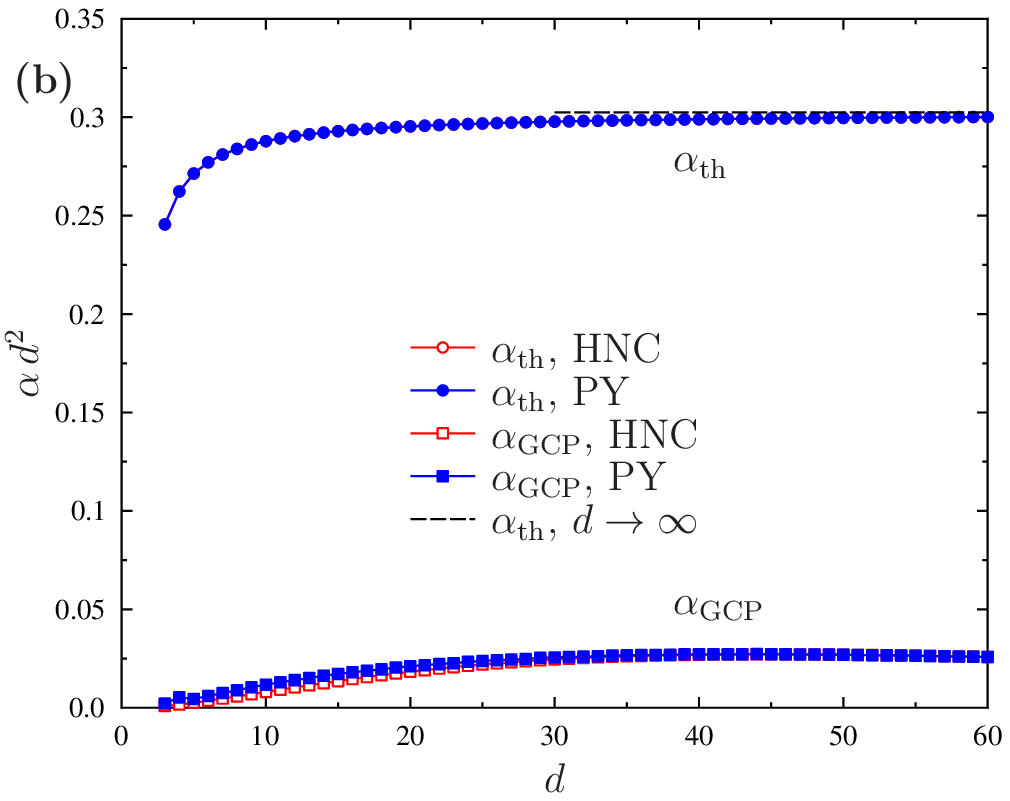}
\caption{(a)~Rescaled threshold $\f_{\rm th}$ and glass close packing $\f_{\rm GCP}$ packing fractions as a function of spatial dimension $d$.
(b)~Rescaled values of the cage radius $\wh \a = \a \, d^2$ at the threshold and GCP points. 
}
\label{fig:2}
\end{figure}

In Figures~\ref{fig:1} and \ref{fig:2} we report results for the transition densities, obtained using the HNC
and the PY approximations as a function of $d$.
The results are naturally expressed as a function of the rescaled packing fraction $\wh\f = 2^d \f/d$, which has a finite
limit for $d\to\io$~\cite{PZ10} at the dynamical transition. We see in Figure~\ref{fig:1} that the rescaled dynamical transition
$2^d \f_{\rm d}/d$ in fact
converges, for both HNC and PY, towards the value of $\wh\f_{\rm d} = 4.8067\ldots$ predicted by the solution in $d\to\io$.
The Kauzmann transition density, instead, grows continuously and for large dimensions we recover the asymptotic behavior
$2^d\f_{\rm K}/d \sim \log d$ predicted in~\cite{PZ10}.
The value of the cage radius at $\f_{\rm d}$ also converges, for large $d$, to the scaling $A \sim d^{-2}$, with 
$\wh A_{\rm d} = A_{\rm d} d^2 \to 0.5766$ for $d\to\io$, as predicted by 
the infinite-dimensional solution. Note that we observe a jump in $A_{\rm K}$ around $d=30$ that is due to the fact that we 
used an asymptotic approximation for the Bessel functions at large distances, because of numerical precision issues with the Bessel package. 
This shows that
$A_{\rm K}$ is very sensitive to numerical precision, however we find that the values of the complexity and of the transition densities
are much more stable numerically.

Interestingly, we find that within the HNC approximation one has $\f_{\rm d} < \f_{\rm K}$ only for $d>10$. For $d\leq 10$, 
the equilibrium complexity $\Si_{\rm eq}$ is always negative. This is unphysical within the mean field picture and we thus conclude
that the HNC approximation does not provide sensible results for low dimensions. The PY results remain physical down to
$d=2$, but we will see later that they are also quite inaccurate.

In Figure~\ref{fig:2} we report the values of packing fraction that delimit the jamming line: the threshold $\f_{\rm th}$
and the glass close packing $\f_{\rm GCP}$. Note that, as shown in Sec.~\ref{sec:jamming}, these quantities depend only
on $g_{\rm liq}(1;\f)$ which is related to the pressure, hence only the equation of state is needed to compute them, and not
the full shape of the pair distribution function. 
As for the equilibrium quantities, in the $d\to\io$ limit we recover the exact results $2^d \f_{\rm th}/d = 6.2581\ldots$ and
$2^d \f_{\rm GCP}/d \sim \log d$~\cite{PZ10}. Similarly, $\wh\a_{\rm th} = \a_{\rm th} \, d^2 \to 0.3024$ for large $d$.
Note that in this case both the HNC and PY approximations provide a region
of positive complexity (a finite jamming line) in all dimensions down to $d=2$. Note also that when $d\to\io$, the threshold density falls
in a region of instability of the RS solution we used here, which leads to an underestimation
of the correct threshold~\cite{CKPUZ13}: probably, the same phenomenon happens in finite $d$.

\subsection{Three dimensional results: thermodynamics}

\begin{figure}[t]
\includegraphics[width=.45\textwidth]{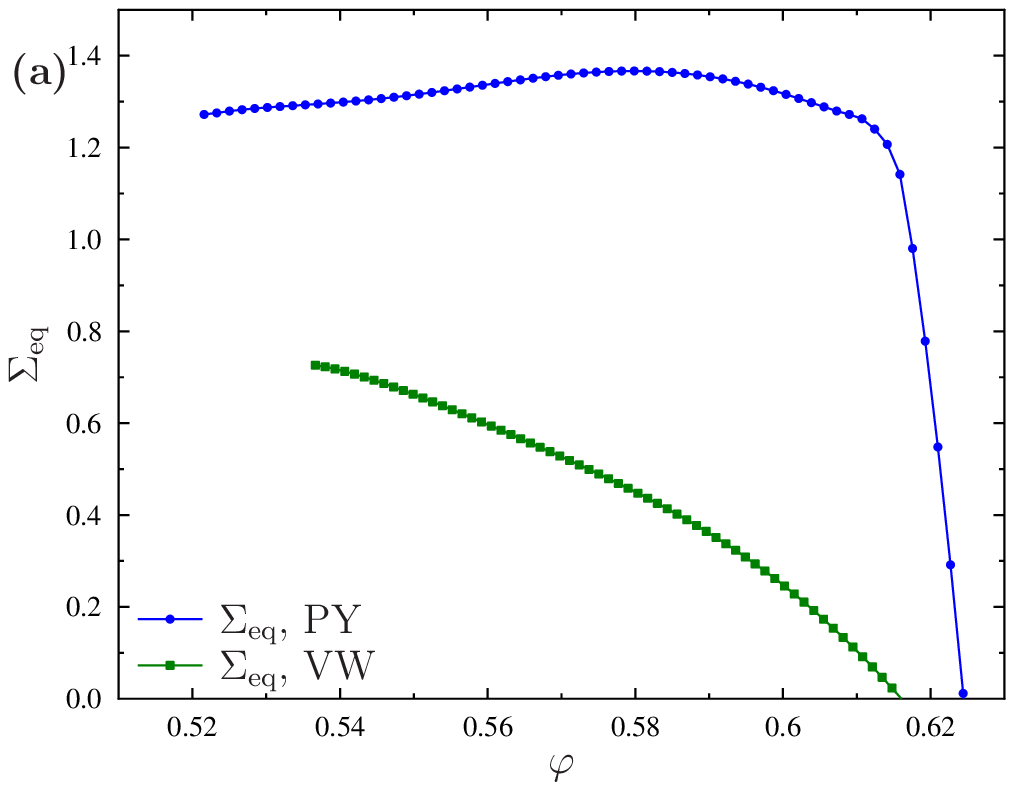}
\includegraphics[width=.45\textwidth]{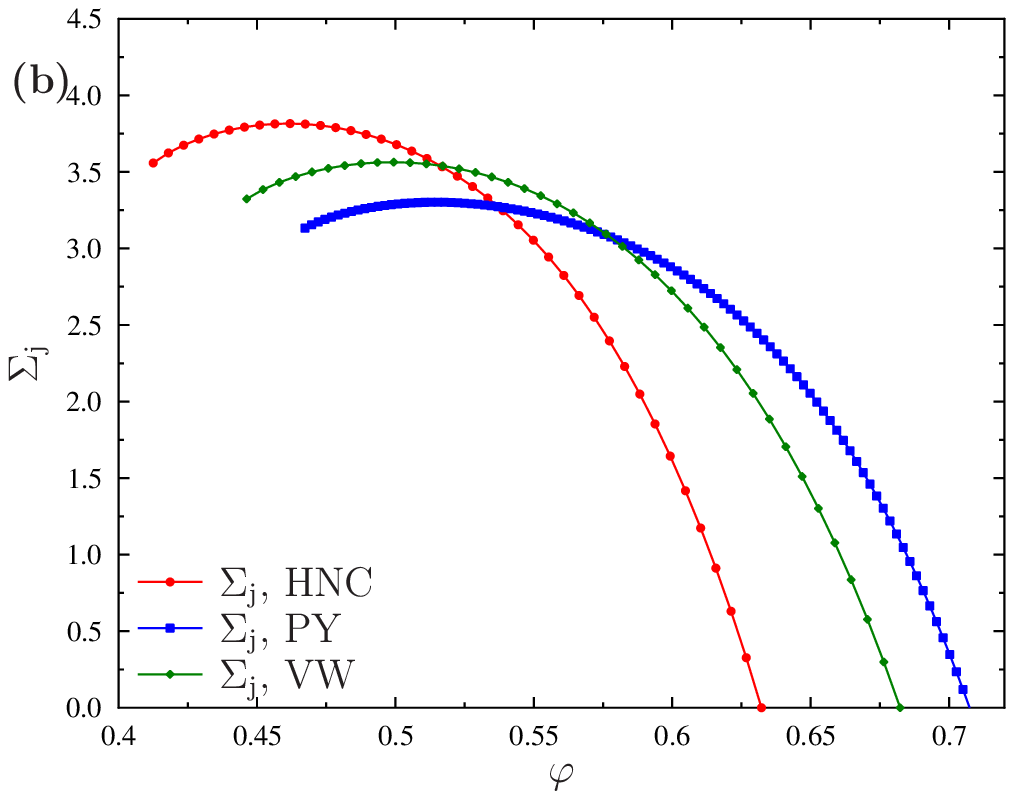}
\caption{
Results for hard spheres in $d=3$. (a)~Equilibrium complexity and
(b)~jamming complexity as a function of packing fraction.
}
\label{fig:3}
\end{figure}

Let us now focus more explicitly on the results in $d=3$. The equilibrium and jamming complexities are reported in Figure~\ref{fig:3},
and the relevant densities in Table~\ref{tab:1}.
Before commenting the results, let us stress that the complexity results from a delicate cancellation between the liquid entropy which
is $\sim 10$ and the glass internal entropy which is also $\sim 10$, the difference between the two being $\sim 1$~\cite[Fig.3]{PZ05}. 
Thus, a relative error of order $10$\% on either the liquid or the glass entropy is enough to obtain a completely meaningless result 
for the complexity. Based on this, we believe that both the PY and HNC closures are unreliable to determine the complexity,
because they have an error precisely of order $10$\%~\cite{Hansen}
 in the relevant density range.
The VW closure is much more accurate and indeed, as we will see,
the results for both the equilibrium and the jamming complexities are reasonable.

For $\Si_{\rm eq}$, we already noticed that the HNC approximation gives a result that is always negative and therefore unphysical;
we can still compute the value of $\f_{\rm d}$ (Table~\ref{tab:1}) but the complexity at $\f_{\rm d}$ is negative.
The PY result has a region of positive complexity. The values of $\f_{\rm d}$ and $\f_{\rm K}$ are reasonable but the shape of
the complexity is quite strange, with a very sharp drop close to $\f_{\rm K}$: this is due to the behavior of the liquid entropy
within the PY approximation, which is quite inaccurate, and we believe that the result is not very reliable.
Because both the PY and HNC approximations do not provide good results in low dimensions, we tested the VW form that gives
very accurate results for the liquid quantities. Using this approximation, we find that the complexity $\Si_{\rm eq}$ has a very reasonable
shape, with a smooth drop that reminds the one obtained numerically~\cite{AF07,BC14}. In particular, the value of $\Si_{\rm eq}$ at $\f_{\rm d}$ is similar
to the one found in~\cite{BC14} on a different system.
Note that the value of $\f_{\rm d}$ is clearly underestimated by all the liquid approximation schemes, a fact that also happens within MCT~\cite{Go09}:
mean field approaches have a tendency to stabilize the glass phase because they neglect 
important relaxation pathways of the liquid phase (e.g. hopping, see~\cite{CJPZ14} for a quantitative estimate of the impact of hopping
on $\f_{\rm d}$). Instead, the value of $\f_{\rm K}$ is close to the one obtained within the small cage approximation~\cite{PZ10},
and is close to the current numerical estimates~\cite{BEPPSBC08} (which however are obtained on polydisperse systems), see Table~\ref{tab:1}.

The jamming complexity $\Si_j$ is the logarithm of the number of jammed states, hence it coincides with the Edwards entropy~\cite{EO89,Ed94}.
All the three approximations (PY, HNC, VW) give a reasonable shape for this quantity. However, PY seems to overestimate the
value of $\f_{\rm GCP}$ (it is very close to the value 0.74 corresponding to the crystalline close packing), while HNC seems to underestimate it (it is below values
of packing fraction 0.64 that are routinely obtained in simulations and experiments). 
Overall, once again 
VW seems to provide the most reliable results. Let us recall that in all cases the RS solution is expected to be unstable
on the jamming line, leading in particular to an underestimation of
the value of $\f_{\rm th}$~\cite{CKPUZ13}.

A comparison of the theoretical results with the best currently available numerical estimates is given in Table~\ref{tab:1}.
Note that no numerical estimate of $\f_{\rm th}$ and $\f_{\rm GCP}$ is available: this is because sampling jammed packings
uniformly (``\`a la Edwards'') is extremely hard~\cite{XFL11,APF14,MSSWF15}. 
Therefore, in numerical simulations and experiments jammed packings are constructed by protocols
that sample packings with non-uniform weights and the final jamming density $\f_{\rm j} \in [\f_{\rm th}, \f_{\rm GCP}]$ can be anywhere on
the jamming line depending on the details of the protocol~\cite{DTS05,SDST06,CBS09,CIPZ11}.
 In the special case where glassy states are followed adiabatically, one can compute the final jamming density using the ``state following''
 construction: this calculation has been done in $d\to\io$~\cite{RUYZ14} and it could be also done using our finite-dimensional approximation
 scheme. We leave this for future work.

\begin{table}
\begin{tabular}{|c|c|c|c|c|}
\hline
Approximations/densities & $\varphi_{\rm d}$ & $\varphi_{\rm K}$ & $\varphi_{\rm th}$ & $\varphi_{\rm GCP}$ \\
\hline
HNC & $0.5596$ & - & $0.4125$ & $0.6322$ \\
PY & $0.5216$ & $0.6244$ & $0.4674$ & $0.7056$ \\
VW & $0.5367$ & $0.6161$ & $0.4463$ & $0.6823$ \\
\hline
\cite{CJPZ14} & $0.5770(5)$ & - & - & - \\
\cite{BEPPSBC08} & 0.590(5) & 0.637(2) & - & - \\
\hline
\end{tabular}
\caption{Theoretical values of the transition densities for hard spheres in $d=3$ (upper part of the table) and corresponding
most recent numerical estimates (lower part of the table). Note that the numerical estimates are for polydisperse systems and
the numbers are quite sensitive to polydispersity (typically, polydispersity increases the value of the transition densities).}
\label{tab:1}
\end{table}

\subsection{Three dimensional results: non-ergodicity factor}

\begin{figure}[t]
\includegraphics[width=.45\textwidth]{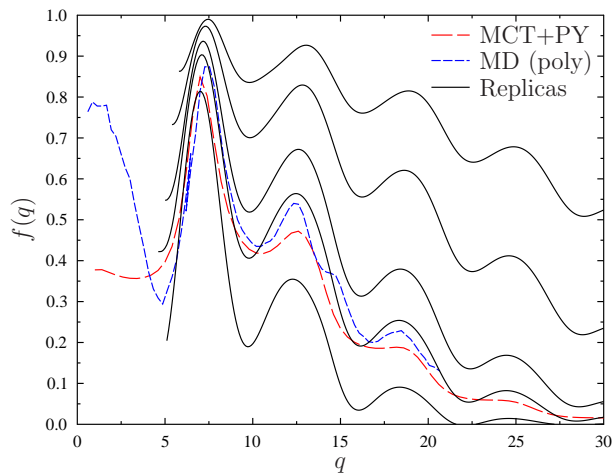}
\caption{
Non-ergodicity factor of hard spheres in $d=3$. Replica results are obtained from the Verlet-Weis approximation
for $\f = 0.537 \ (=\f_{\rm d}), 0.548, 0.560, 0.586, 0.616 \ (=\f_{\rm K})$ (black lines, from bottom to top).
MCT results using the PY structure for the liquid at $\f=\f_{\rm MCT}^{\rm PY} = 0.516$~\cite{FFGMS97} are reported
as a red long-dashed line.
Numerical results from MD~\cite{WPFV10} for polydisperse hard spheres at $\f=0.585$, which is estimated to
be close to $\f_{\rm d}$ for this system, are reported as a blue dashed line.
}
\label{fig:fq}
\end{figure}

In Figure~\ref{fig:fq} we report the results for the non-ergodicity factor $f(q)$ using the Verlet-Weis approximation,
at $m=1$ corresponding to the equilibrium system.
Results are reported at several densities ranging from $\f_{\rm d}$ to $\f_{\rm K}$.
As discussed in Section~\ref{sec:fqnum}, the numerical calculation is quite heavy and it is affected by spurious effects
related to the various cutoffs for $q\lesssim 5$, so we do not report the data in that regime. We checked that the data
we report are independent of the cutoffs.

For all densities $f(q)$ has the expected shape and $f(q)$ increases towards 1 for all $q$ upon increasing densities,
which is reasonable because the glass becomes more and more stable. At $\f=\f_{\rm d}$, we find that $f(q)$ becomes
slightly negative around $q\approx 22$, which might be either a numerical artifact (although we could not find its
origin) or an unphysical feature due to the approximations involved in the theory. 

In the figure, we compare the results with the MCT prediction, obtained solving the MCT equations using the PY
approximation as input, for hard spheres in $d=3$~\cite{FFGMS97}. 
We also compare with numerical data obtained in~\cite{WPFV10} using Molecular Dynamics (MD) 
for a polydisperse system of hard spheres. The agreement of our results with MCT and with numerics is rather good,
although the VW approximation at $\f_{\rm d}$ gives more pronounced oscillations with in particular
a lower value of $f(q)$ at the local minima.
One should take into account the fact that determining $\f_{\rm d}$ precisely in numerical simulations is not an easy
task and there is always some ambiguity, see e.g.~\cite{CJPZ14}. Thus, it is possible that the numerical result 
effectively corresponds to $\f > \f_{\rm d}$.
A more detailed comparison requires repeating
the replica computation for a polydisperse system, following~\cite{BCPZ09}.

\section{Conclusions}

In this paper we presented an approximation scheme to compute quantitatively the main observables in glassy systems, within
a mean field approach. Our method is based on an expansion around the limit $d\to\io$ and can, in principle, be systematically
improved by adding more terms to the expansion. This should be done by adding more diagrams, possibly containing many-body 
replica interactions~\cite{PZ10}, and also perturbing around the Gaussian approximation for $\r(\bar x)$ and the factorised approximation
for $g(\bar x,\bar y)$, Eqs.~\eqref{gprod} and \eqref{rrho}.

Here we considered the simplest approximation scheme.
The resulting equations, derived in Section~\ref{sec:derivation}, take as input equilibrium properties of the liquid phase and give
as output the properties of the glass and of jamming. They can be used for any interaction potential and in any spatial dimension.
Although in this paper we focused on monodisperse systems and on the replica symmetric scheme, the equations are easily generalized
to take into account polydispersity~\cite{BCPZ09} and additional steps of replica symmetry breaking~\cite{CKPUZ13}.
Our equations are conceptually similar to MCT ones, and their solution requires a comparable computational power. 
The advantage is that we have access to many physical quantities, both in the supercooled liquid regime and deep in the glass
phase, including jamming. 

Note that in the approximation scheme used in this work, the only required input (at least to compute the dynamical transition point)
is the liquid pair correlation $g_{\rm liq}(r)$. Thus, for potentials having the same $g_{\rm liq}(r)$ the theory will predict the same
dynamical transition point. It has been shown in~\cite{BT10} that for some potentials this is not correct. This finding suggests that three-body,
and higher order, liquid correlations might play an important role in glassy behavior for some systems. Although here we neglected these correlations,
they can be introduced by adding more terms in our systematic expansion~\cite{PZ10}.

In this paper we only focused on $d$-dimensional hard spheres and we have shown that in the limit $d\to\io$ we recover the exact solution,
and that in low dimensions we obtain reasonable numbers for the transition densities and the complexity. 
Moreover, the result for the non-ergodicity factor $f(q)$ is quite close to the one obtained from MCT.
However, all these results are
very sensitive to the quality of the approximation used for the equilibrium liquid quantities. Therefore, in order to make a precise comparison
between theory and simulation or experiment, one should {\it (i)} choose a system, {\it (ii)} find an accurate way of computing the liquid quantities
for that specific system, and {\it (iii)} solve our equations and compare the results with the simulation/experiment. Given the large amount
of data available for several different systems, we hope that this will be done in the future.

\acknowledgments

This study was motivated by discussions with Patrick Charbonneau,
Giorgio Parisi and Matteo Rosati: we are very grateful to them for the inspiration and for
useful remarks. We would like to thank Atushi Ikeda for sharing his code to solve liquid theory in $d>3$, 
Andr\'es Santos for sharing a code to solve exactly the Percus-Yevick equation in $d>3$,
and Matthias Fuchs and Antonio Puertas for sharing the MCT and MD result for $f(q)$.
We also thank Grzegorz Szamel and Walter Kob for very useful discussions.

\vskip-15pt

\bibliographystyle{mioaps}
\bibliography{HS}

\end{document}